\documentclass{aa}  

\usepackage{graphicx}
\usepackage{txfonts}

\usepackage{siunitx}
\usepackage{hyperref}
\usepackage{lscape}

\newcommand{\gaia}{\textit{Gaia\,}}
\defcitealias{gaia_collaboration_gaia_2016}{Gaia Collaboration, Brown et al. (2016)}

\newcommand{\orcid}[1]{\protect\href{https://orcid.org/#1}{\protect\includegraphics[width=8pt]{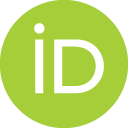}}}

\makeatletter
\renewcommand*\aa@pageof{, page \thepage{} of \pageref*{LastPage}}
\makeatother

\begin{document}

   \title{Near-infrared spectroscopic characterisation of Gaia ultra-cool dwarf candidates
   \thanks{Based on observations made with the ESO New Technology Telescope at the La Silla Observatory under programme 106.214E.001 and 108.22G4.001}}

   \subtitle{Spectral types and peculiarities}

   \author{T.~Ravinet\orcid{0000-0001-8652-2835}
          \inst{1}
          \and
          C.~Reylé\orcid{0000-0003-2258-2403}
          \inst{1}
          \and
          N.~Lagarde\orcid{0000-0003-0108-3859}
          \inst{2}
          \and
          A.~Burgasser\orcid{0000-0002-6523-9536}
          \inst{3}
          \and
          R.\,L.~Smart
          \inst{4}
          \and
          W.\,H.~Moya
          \inst{5, 6}
          \and 
          F.~Marocco
          \inst{7}
          \and
          R.-D.~Scholz
          \inst{8}
          \and
          W.\,J.~Cooper\orcid{0000-0003-3501-8967}
          \inst{9,4}
          \and
          K.\,L.~Cruz
          \inst{10}
          J.\,G.~Fern\'andez-Trincado
          \inst{11}
          \and
          D.~Homeier
          \inst{12}
          \and
          L.\,M.~Sarro
          \inst{13}
          }

   \institute{Universit\'e de Franche-Comt\'e, CNRS UMR6213, Institut UTINAM, OSU THETA Franche-Comt\'e-Bourgogne, Observatoire de Besan\c con, BP 1615, 25010 Besan\c con Cedex, France
         \and
         Laboratoire d’Astrophysique de Bordeaux, Université Bordeaux, CNRS, B18N, Allée Geoffroy Saint-Hilaire, 33615 Pessac, France
         \and
        Department of Astronomy \& Astrophysics, University of California San Diego, La Jolla, CA 92093, USA
        \and
        Istituto Nazionale di Astrofisica, Osservatorio Astrofisico di Torino, Strada Osservatorio 20, I-10025 Pino Torinese, Italy
        \and
        Instituto Multidisciplinario de Investigaci\'on y Postgrado, Universidad de La Serena, Ra\'ul Bitr\'an 1305, La Serena, Chile
        \and
        Departamento de Astronom\'ia, Universidad de La Serena, Av. Cisternas 1200, La Serena, Chile
        \and 
        IPAC, Mail Code 100-22, Caltech, 1200 E. California Boulevard, Pasadena, CA 91125, USA
        \and
        Leibniz-Institut für Astrophysik Potsdam, An der Sternwarte 16, 14482 Potsdam, Germany
        \and
        Centre for Astrophysics Research, University of Hertfordshire, Hatfield, Hertfordshire, AL10 9AB, UK
        \and 
        Hunter College, Physics and Astronomy, 695 Park Avenue, New York, NY 10065, USA
        \and
        Instituto de Astronom\'ia, Universidad Cat\'olica del Norte, Av. Angamos 0610, Antofagasta, Chile
        \and
        Aperio Software Ltd., Insight House, Riverside Business Park, Stoney Common Road, Stansted, Essex, CM24 8PL, UK
        \and 
        Dpto. de Inteligencia Artificial, UNED, c/ Juan del Rosal 16, 28040 Madrid, Spain
             }

   \date{Submitted 13/09/2023, Revised 26/01/2024, Accepted 30/01/2024}

  \abstract
     {  The local census of very low-mass stars and brown dwarfs is crucial to improving our understanding of the stellar-substellar transition and their formation history. These objects, known as ultra-cool dwarfs (UCDs), are essential targets for searches of potentially habitable planets. However, their detection poses a challenge because of their low luminosity. The \gaia survey has identified numerous new UCD candidates thanks to its large survey and precise astrometry.}
   {We aim to characterise 60 UCD candidates detected by \gaia in the solar neighbourhood with a spectroscopic follow-up to confirm that they are UCDs, as well as to identify peculiarities.}
   {We acquired the near-infrared (NIR) spectra of 60 objects using the SOFI spectrograph between 0.93 and 2.5 microns (R\(\sim600\)). We identified their spectral types using a template-matching method. Their binarity is studied using astrometry and spectral features.}
   {We confirm that 60 objects in the sample have ultra-cool dwarf spectral types close to those expected from astrometry. Their NIR spectra reveal that seven objects could host an unresolved coolest companion and seven UCDs share the same proper motions as other stars. The characterisation of these UCDs is part of a coordinated effort to improve our understanding of the Solar neighbourhood. }
  {}


    \titlerunning{Gaia ultra-cool dwarfs spectral types}
    \authorrunning{T. Ravinet et al.} 

   \maketitle
%

\section{Introduction}

    M dwarfs (\(\lesssim0.6~\mathrm{M}_\odot\)) are the most prominent stars in the Galaxy and account for $\sim60\%$ of the stellar budget in the solar neighbourhood \citep{bochanski_luminosity_2010, reid_low-mass_1997, reyle_10_2021}. Objects with spectral types later than M7, with a temperature lower than \(2800\,\si{K}\) \citep[e.g. ][]{rajpurohit_effective_2013} have been defined as ultra-cool dwarfs (hereafter, UCDs) by \cite{kirkpatrick_ultra-cool_1997}, and contains the less massive stars, as well as brown dwarfs, encompassing the stellar-substellar transition.

    The study of UCDs atmospheres is complex, due to their low temperature, necessitating the inclusion of the effects of dust and condensate in their surface layers \citep{tsuji_dust_1996, allard_limiting_2001, allard_progress_2013}. In addition, the possible formation of clouds \citep{ackerman_precipitating_2001, saumon_evolution_2008} and strong vertical mixing due to eddy diffusion \citep{noll_detection_1997, geballe_spectroscopic_2009, phillips_new_2020} complicates our understanding.
    To describe the stellar-substellar transition, at which the hydrogen is no longer burning in a stable way, new equations of state have been developed \citep{fernandes_evolutionary_2019, chabrier_new_2021, chabrier_impact_2023}. These descriptions of different mechanics occurring inside cool objects, associated with atmosphere models, allow for the evolution of UCDs to be modelled and for estimations of their masses and ages to be obtained.

    Observations of UCDs spanning the variety of their spectral types, ages, and atmospheres illustrate model limitations on the overall shape of spectra, and further constrain them \citep{beiler_l-band_2023}. 
    Large surveys and spectroscopic follow-ups of red, cold, and low-luminosity objects have provided a sufficient sample to statistically study UCDs. Luminosity functions have been built to determine the density of UCDs in the solar neighbourhood \citep{cruz_meeting_2007, reyle_ultracool-field_2010, bochanski_luminosity_2010}. Moreover, indicators of spectral binarity \citep{burgasser_spex_2010, burgasser_hubble_2011, bardalez_gagliuffi_spex_2014} permit to estimate the fraction of unresolved objects in surveys and to correct luminosity functions for that effect \citep{bardalez_gagliuffi_ultracool_2019}.

    However, the identification of UCDs requires the detection capability of faint, elusive objects that emit most of their light in the near-infrared (NIR). \cite{bardalez_gagliuffi_ultracool_2019}, using the UCD density found up to 20~pc from the Sun, estimated that the UCD census is incomplete within 25~pc of the Sun, with \(69-80\%\) of M7-L5 catalogued. The \gaia space mission \citepalias{gaia_collaboration_gaia_2016} offers a new way to discover UCD candidates based on astrometry and photometry. Using a theoretical approach, \cite{sarro_properties_2013} estimated that more than 40 000 objects should be observed by \gaia. Thanks to its complete sky coverage, the satellite permits to unveil the optical emissions of UCDs down to a  G-magnitude of~\(\sim 20.5\). The unprecedented precision of the parallaxes obtained by the satellite allows for the selection of the coolest objects based on their locus in colour-absolute magnitude diagrams, avoiding the contamination from giant stars.
    With the \gaia Data Release 2, \cite{reyle_new_2018} compiled a set of 14~200 UCD candidates, with more than a thousand of them within 50~pc of the Sun, which was completed by the candidates found by \cite{smart_gaia_2019, scholz_new_2020}. Using the \gaia Data Release 3, \cite{smart_gaia_2021, sarro_ultracool_2023} acquired new candidates and their parallaxes. These new discoveries are now being studied via spectroscopic follow-ups.
    
    In this study, we acquire the spectra of UCD candidates to confirm their spectral type. These new spectra also allow us to improve our understanding of the physics and atmospheric processes occurring in these cool objects and to search for unresolved binaries. Here, we present the spectra of 60 UCD candidates, acquired with the SOFI spectrograph at the New Technology Telescope (NTT) in la Silla observatory. In Sect. 2, we describe the sample, the acquisition, and reduction of the spectra of these candidates. In Sect. 3, we discuss the method used to obtain spectral types of the UCD candidates and confirm that most of the objects are UCDs. In Sect. 4, we discuss peculiar objects, common parallax and proper motion binaries and spectral binaries present in our sample. We present our conclusions and give our perspectives on the use of this sample in Sect. 5.

\section{Sample description, acquisition, and reduction}
    \subsection{Target selection}

    Numerous \gaia UCD candidates have been revealed by \cite{reyle_new_2018, smart_gaia_2019} and \cite{scholz_new_2020}. The spectroscopic confirmation of the closest ones, corresponding to 228 targets, is driven by a coordinated effort from different observatories and instruments. Our observations monitored 60 UCD candidates, which is an initial phase towards the full publication of this spectroscopic study. The 13 candidates that are closer than 25 parsec are contained in the complete \textit{Catalogue of Nearby Stars 5} \citep{golovin_fifth_2023}.
    
\begin{figure*}[!hbt]
        \centering
        \includegraphics[width=\linewidth]{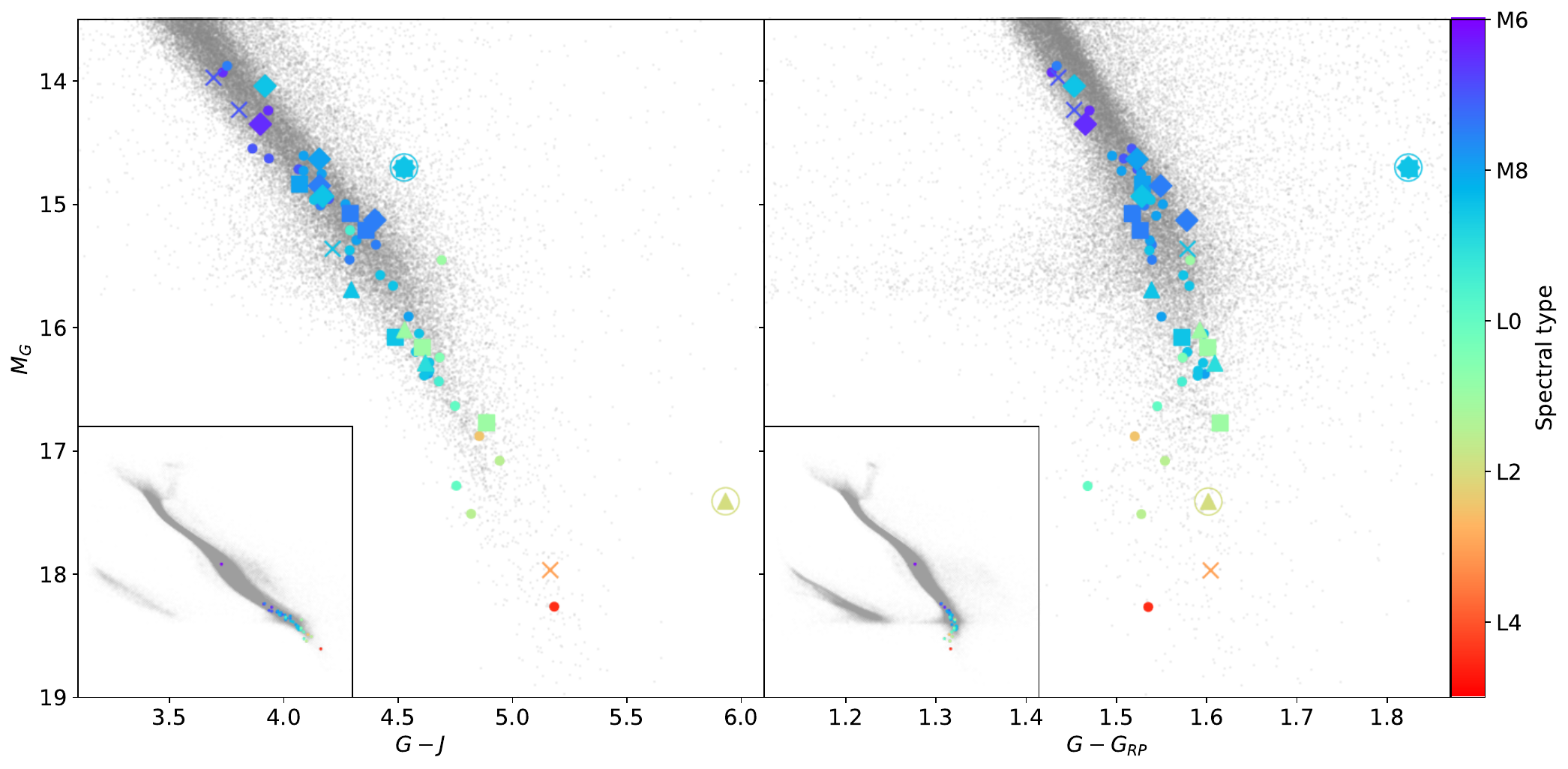}
        \caption{Observed UCD candidates on a zoomed \gaia \(M_G\) -  \(G-J\) and \(G-G_{RP}\) diagrams: UCDs are coloured by their obtained spectral type, as explained in Sect. 3. 
        The two circled UCDs have peculiar colours (see text).
        Subdwarfs candidates are represented by crosses (see Sect. \ref{sec:potsubd}). Peculiar objects are represented by triangles (Sect. \ref{sec:peculiars}).
        The UCDs with common parallaxes and proper motions with another system are represented by diamonds (see Sect. \ref{sec:commonprop}) and those in (potentially) unresolved binary systems by squares (Sect. \ref{sec:specbin}). 
        The grey points are the sources from the \gaia Catalogue of Nearby Stars, with a RUWE \(<1.4\).
        The insets illustrate the locus of our sample on entire \gaia Catalogue of Nearby stars colour-absolute magnitude diagrams. }
        \label{fig:sofiGrpCAMD}
    \end{figure*}

    The 60 potential UCDs of this sample were selected as visible during the timeframe and location of our observations (see Sect. \ref{sec:acquisition}). They are drawn from the 50~pc UCD candidates sample or because they have common parallaxes and proper motions with other stars (Sect. \ref{sec:commonprop}). They are shown on the colour-absolute-magnitude diagrams in Fig. \ref{fig:sofiGrpCAMD}, superimposed with the \gaia Catalogue of Nearby Stars \citep[Gaia Collaboration, ][]{smart_gaia_2021}. Their names, \gaia identifiers, parallaxes, proper motions, and tangential velocities are listed in Table  \ref{tab:sofiobservations}.

   Two objects with peculiar colours are highlighted on Fig. \ref{fig:sofiGrpCAMD}. 
   J0347+0417 has a redder \(G - G_{RP}\) colour than the rest of the sequence. This behaviour is not present in its 2MASS colours. We find that it has a relatively high \gaia RUWE\footnote{Re-normalised unit weight error ; quantifies the goodness-of-fit of the \gaia astrometric solution} (1.27) and a positive \textit{IPD\_ frac\_multi\_peak}\footnote{Percentage of windows, for which the IPD algorithm has identified a double peak, meaning that the detection may be a visually resolved double star} (5), which can be correlated with partially unresolved binarity. The object is also very tight in the sky with \gaia DR3 3271777035212786944, with which it shares common parallaxes and proper motions (see Sect. \ref{sec:commonprop}), a source that was not resolved in 2MASS images. Further details on this source are given in Sect. \ref{sec:specbin}, explaining the red \gaia colours. 
   J0412-0734 is very red in \(G-J\), due to a very poor photometry in the J-band (\(U\) quality flag in the 2MASS survey). 
   
    \subsection{Spectra acquisition}
    \label{sec:acquisition}
    
    Observations were carried in remote mode, during three nights in December 2020 and three nights in March 2022, under the programmes 106.214E.001 and 108.22G4.001, with the SOFI spectrograph  \citep{moorwood_sofi_1998} located on the New Technology Telescope in la Silla. The spectrograph was set with a \(1''\) slit and in a low-resolution mode, with the blue and red grisms, allowing to observe the \(0.90 - 2.53 \si{\micro\meter}\) range with a resolution \(R\sim 600\). Objects were observed in an AB-BA pattern, dithering around the slit. Exposure times varied from 20 seconds to 15 minutes for the blue arm and from 20 seconds to an hour for the red arm, depending on the J and Ks magnitudes of the observed objects.
Telluric standard stars were regularly observed during the night and were drawn from the ESO List of InfraRed Telluric Standards\footnote{\url{https://www.eso.org/sci/observing/tools/standards/IRstandards/Standards_list.html}}, namely, F and G stars.
    Standard calibration images were also captured.
    
    \subsection{Reduction}

    The 2D spectra and calibration images are cross-talk corrected following the SOFI user manual \footnote{\url{https://www.eso.org/sci/facilities/lasilla/instruments/sofi/doc/manual/sofiman_2p50.pdf}}. Flats were combined and averaged over each three-nights observation periods, and then applied to observations images. The wavelength correction is applied using a 5-degree polynomial, through IRAF/pyraf tasks \textit{identify}, \textit{reidentify}, and \textit{fitcoords}. The sky emission in the NIR was removed by subtracting the A-B images along the slit and further cleaned out by using a sky residuals estimate obtained by a rolling average along the wavelength axis.
    The spectra and uncertainties were extracted from images with the IRAF \textit{apall} task in the \textit{onedspec} context. The telluric corrections and flux calibrations were done using packages affiliated with \textit{pyraf} and \textit{Astropy}, as well as \textit{SPLAT} (see Sect. \ref{sec:spec_class_method}). Telluric corrections and flux calibrations make use of the \cite{pickles_stellar_1998} stellar library\footnote{\url{https://www.eso.org/sci/observing/tools/standards/IR_spectral_library.html}}, with model spectra with the closest spectral types to the observed standards. This reduction procedure is drawn from the one used by the PESSTO survey \citep{smartt_pessto_2015}. The signal-to-noise ratios (S/Ns) of the resulting spectra vary from \(\sim 110\) to \(\sim 750\) between objects and the average is \(\sim 300\). The reduced spectra have been made publicly available\footnote{Through the Vizier service at CDS}.
    

\section{Spectral classification}

    \subsection{Methods}
    \label{sec:spec_class_method}

    To derive the spectral-types of our sample of UCD candidates, we used a template-matching method. Using the  \textit{SpeX Prism Library Analysis Toolkit} \citep[SPLAT,][]{burgasser_spex_2017}, we query the SpeX Prism Library (SPL) for a set of spectra of objects with already-assigned spectral types from early M to late L, with a S/N greater than 30 and that have not been identified as spectral binaries. The resulting set contains 670 low-resolution spectra (\(R\sim 120\) for most of the objects). It includes reference UCD spectra, such as the ones defined by \cite{kirkpatrick_discoveries_2010}, but also peculiar ones: low and very-low gravity objects revealed by \cite{cruz_young_2009}, low-metallicity subdwarfs \citep{lodieu_metallicity_2019}, or unusually blue and red M/L dwarfs \citep{kirkpatrick_discoveries_2010}. 

    Each SOFI UCD candidate spectrum \(C[\lambda]\) was smoothed to a lower resolution using a Gaussian window, in order to reach \(R\sim 200\). Then, template spectra \(T[\lambda]\) issued from the SPL were linearly interpolated to a common wavelength scale and compared to \(C[\lambda]\) using a weighted \(\chi^2\) statistic, 
    \[\chi_T^2 \equiv \sum_{\{\lambda\}} w[\lambda]\left[ \frac{C[\lambda] - \alpha T[\lambda]}{\sqrt{\sigma_C[\lambda]^2+\sigma_T[\lambda]^2}} \right]^2,\]
    following the method proposed by \cite{cushing_atmospheric_2008}. We adopt a vector of weights \(w[\lambda]\) where each pixel is weighted by its spectral size (\(w_i\propto \Delta\lambda_i\)) to avoid a bias towards the blue region of the spectra, where the spectral sampling is greater than in the red regions \citep{cushing_atmospheric_2008}. Also, \(\alpha\) is a scaling factor minimising the \(\chi^2\) between a template and an observed spectrum, while \(\sigma_C[\lambda]\) and \(\sigma_T[\lambda]\) are the noise spectra of SOFI observations and of templates, respectively. The S/N of the templates being much lower than the one of SOFI observations, it is essentially the one that is taken into account during the analysis. Following \cite{bardalez_gagliuffi_ultracool_2019}, the template comparisons are made on the \(0.95 - 1.35 \si{\micro\meter} \), \(1.45 - 1.8 \si{\micro\meter,}\) and \(2.0 - 2.35 \si{\micro\meter}\) regions of spectra, to avoid strong telluric absorption wavelength ranges. Then, the spectral type of UCD candidates are derived by averaging the spectral types of the best-matching templates, weighted by their \(\chi_T^2\) statistic. The best-matching templates are selected as the ones with \(\chi^2_T \leq 2 \chi^2_\mathrm{min} \), where \(\chi^2_\mathrm{min} \) is the minimal statistic found in the template list. This procedure ensures that we may avoid wrongly classifying templates in the template sample, in which, for example, some M8 are borderline objects with features of M7 or M9. The \(2\chi^2_\mathrm{min}\) threshold retrieves about ten templates, which is sufficient to filter such misclassifications. We verified that with a higher threshold, the impact on the determined spectral type is of at most 0.5 spectral type. However, the best fits with peculiar templates will be missed when averaging on a larger sample dominated by classical templates.
    For comparison purposes, we also perform the template-matching analysis using only the blue (\(0.95-1.6\mu m\)) and red (\(1.6-2.35\mu m\)) ends of candidates spectra, separately — and excluding the telluric regions. We then visually verified the template-matches to confirm the UCD candidates spectral-types, using the blue or red spectral-types in case of disagreement between observations and full-templates matches. This visual inspection also permits to verify if the best-matching spectral type is peculiar.

    The spectral types found using only blue or red part of spectra are close to the ones using the entire spectrum. Using the blue part alone leads to spectral types on average 0.2 subtypes later, with a standard deviation of 0.5 subtypes around that average. Using the red part alone leads to on average very close results, with no subtype offset, but that are more dispersed, with a standard deviation of 0.8. This (slightly) larger dispersion can be explained by the relatively small number of features in the H and Ks bands of  UCDs, which do not change much across close spectral types, compared to the various molecules and absorption lines appearing in the J-band, as shown in Figures \ref{fig:sofi-spec-1}-\ref{fig:sofi-spec-pec} of Appendix \ref{app:sofi-spectra}. These longer wavelengths are an interesting area of study, as they are more sensitive to metallicity because of collision-induced absorption of molecular hydrogen at low temperatures.
    

\subsection{Results}
\label{sec:results}

\begin{figure}
    \centering
    \includegraphics[width=\linewidth]{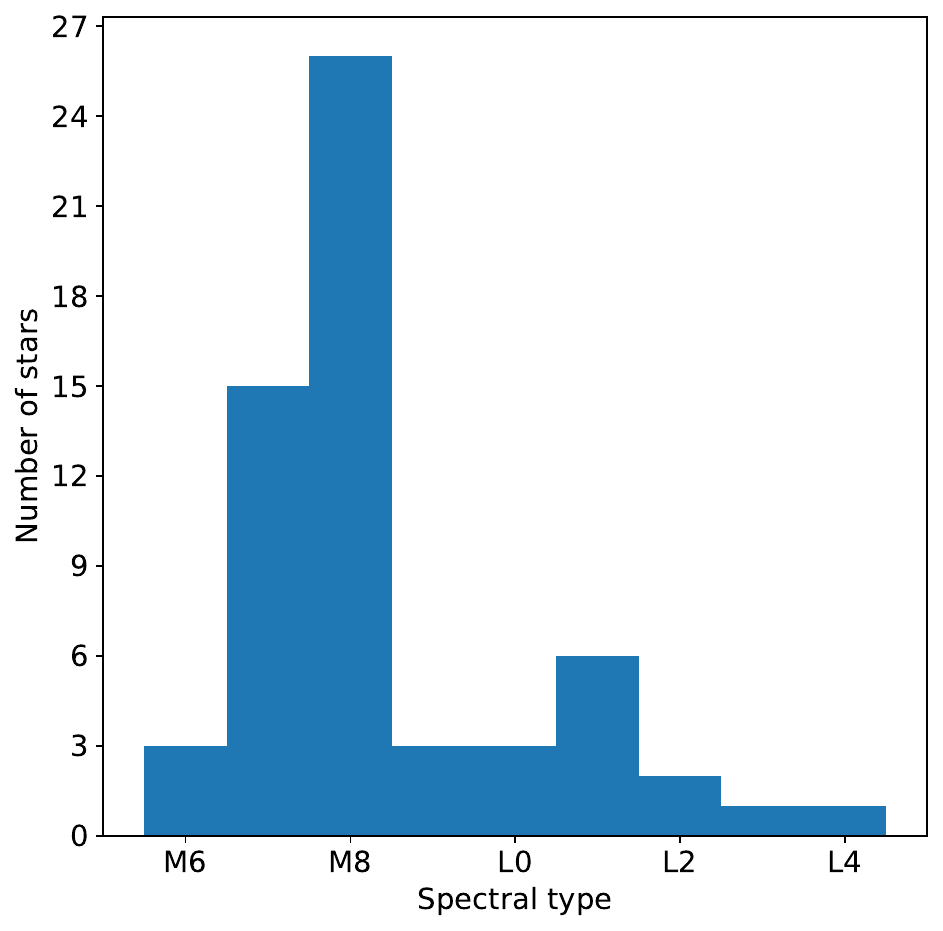}
    \caption{Distribution of the spectral types of our sample.}
    \label{fig:histSOFI}
\end{figure}

 In Table \ref{tab:sofispectral}, we give the names, \gaia DR3 IDs, magnitudes, and spectral types of our UCD candidates samples.  We show their spectra on figures in Appendix \ref{app:sofi-spectra}. We also show the spectra of J0721-3104 (M3), and J2349-2627 (M5), which are binary companions to observed UCDs (see \ref{sec:commonprop}).

 We present on Fig. \ref{fig:histSOFI} the distribution of the spectral types assigned to the objects of our sample. We confirm that all candidates have spectral types later than M7, except three. J0721-3105, J0958-5344 and J1845-2535 are M6.5 stars according to their near-infrared spectra. This spectral type is compatible with these M-dwarfs being UCDs, given the uncertainties of the classification method. The latest spectral-type object in the sample is a L4.5.
The spectral type of six objects have already been studied :
 \begin{itemize}
     \item J0309-1354 is a peculiar M8.5 that was identified as an unsure M6 in the optical range by \cite{kirkpatrick_allwise_2016}. Its peculiarities are discussed in the next section. 
     
     \item J0052-2705 was studied in the optical by \cite{liebert_spectrophotometry_1982} and identified as a peculiar M7.5, and we identify it as having a spectral type of M8.5 in the NIR. 
 
    \item We find that J0703+0711, a M8, was identified with the same spectral type in the optical by \cite{cabello_independent_2019}. 
 
    \item J0109-0343 was identified as a M9 in the optical by \cite{reid_meeting_2003}, and we find it is a L0.5. 
 
    \item J0412-0734 is identified as a peculiar L2 by \cite{kirkpatrick_field_2021} in the NIR. We identify it as having the same spectral type and peculiarity, which we discuss in the next section.
 
    \item J0155+0950 is found to be a L4.5,compatible with the L4 classification found by \cite{burgasser_spex_2010}.
  \end{itemize}

  The published spectral types obtained from NIR spectra are in accordance with our measurements and the differences originate from using different methods and template samples to find spectral types. Moreover, differences between spectral types obtained from NIR and optical spectra are common, as these wavelength ranges probe different regions of the UCDs atmospheres \citep{kirkpatrick_discoveries_2010}.

\begin{figure}
    \centering
    \includegraphics[width=\linewidth]{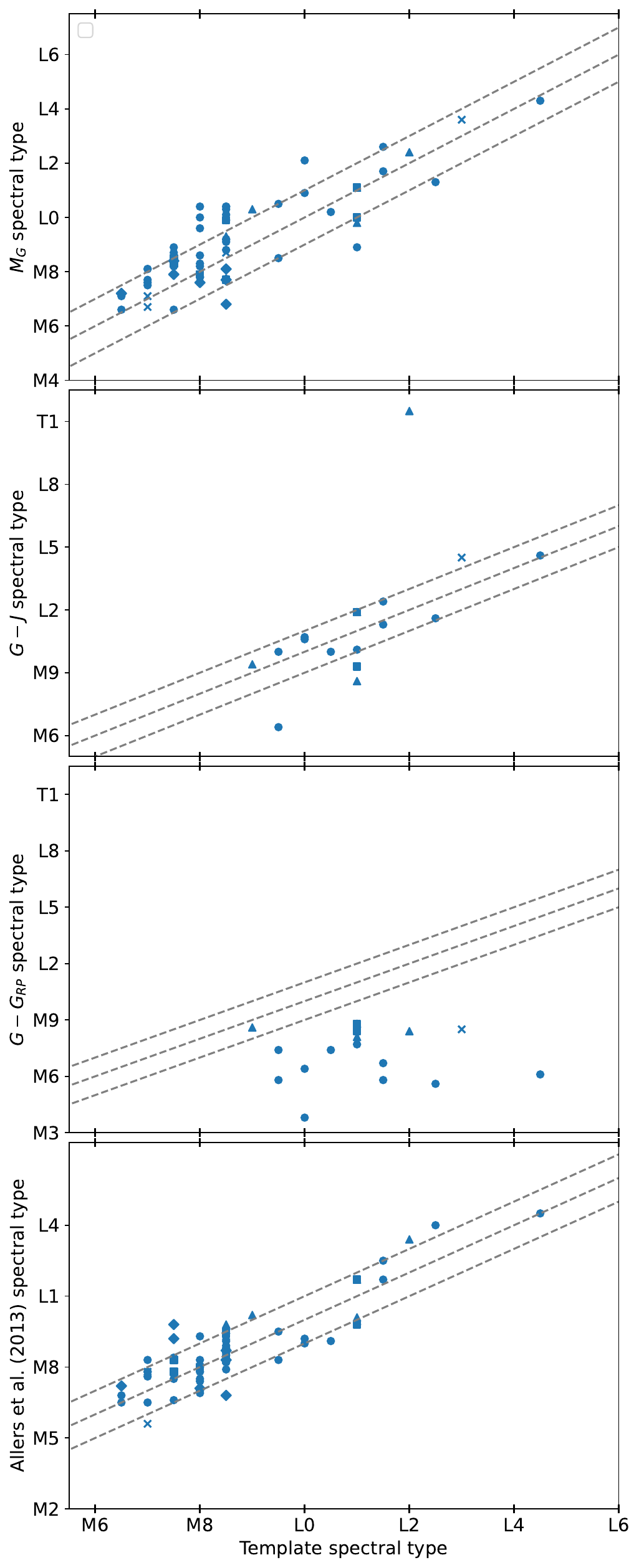}
    \caption{Comparison between the spectral types obtained with \cite{reyle_new_2018} \(M_G\), \(G-J\), \(G-G_{RP}\) photometric spectral types (first, second, and third panels from top, respectively) and \cite{allers_near-infrared_2013} spectral indices (bottom panel). Dashed lines show the identity lines plus/minus one spectral type. Symbols are the same as those used in Fig. \ref{fig:sofiGrpCAMD}.}
    \label{fig:sptrel}
\end{figure}

 We compare our results with various methods to derive spectral types, either from photometry, spectral indices and template-fitting methods. 
 In the top panel of Fig. \ref{fig:sptrel}, we compare the spectroscopic spectral type to the photometric (\(M_G\)) type, published by \cite{reyle_new_2018} and used to select UCD candidates in \gaia DR2. We find that the published spectral types relatively close to  our results, with photometric types slightly later than the spectroscopic ones, with an on-average difference of \(+0.4\) subtypes and a standard deviation of 1.0 subtype. 

 Additionally, \cite{reyle_new_2018} published absolute-magnitude and colour-spectral type relations, which provide photometric spectral types given the astrometry of UCD candidates. Using these, we compared the photometric types obtained from the 2MASS \(J,H,Ks\) absolute magnitudes to our results: The \(M_J\) photometric types differ by \(+0.4\pm1.1\) types compared to our spectroscopic types, while \(M_H\) and \(M_{Ks}\) types are later by \(0.6\pm1.1\) and \(0.8\pm 1.2\).
 We also show in the second panel of Fig. \ref{fig:sptrel} the photometric types based on the \(G-J\) colour, which are (on average) \(0.3\pm2.7\) later than the spectroscopic ones. The outlier point, with a photometric spectral type of T1.5 is J2349-2627, a L2.5 found in a binary system, discussed in Sect. \ref{sec:peculiars_and_bins}. The \(G-G_{RP}\) colour estimate photometric types with an offset of \(-4 \pm 2.0\) compared to the spectroscopic ones (third panel). This large difference in spectral type determination is explained by the spectral energy distribution of UCDs, which are predominantly radiating in the NIR. As can be seen in Fig. \ref{fig:sofiGrpCAMD}, the colour of the M and early L UCDs does not vary much with spectral type, resulting in a poor spectral type determination with this colour.
We thus note that absolute magnitude relations perform better than colours ones, and should be preferred if an accurate estimation of distance is available.
 
 Additionally, \cite{allers_near-infrared_2013} used four gravity-insensitive indices issued from \cite{mclean_nirspec_2003, slesnick_spectroscopically_2004} and \cite{allers_characterizing_2007}, all based on \(H_2 O\) features, to derive spectral types from NIR spectra.  We compare the results given by the average of the indices to the spectral type of the template-fitting method. We find that the spectral types obtained with this method are similar to the ones found in our study, with a small offset and dispersion of \(0.13 \pm 0.90\), as visible in the lower panel of Fig. \ref{fig:sptrel}. 

 \subsection{Effective temperatures}
 \label{sec:efftemp}

 We derived effective temperatures for the UCDs by comparing them with synthetic spectra. Our approach relies on maximising the likelihood \(\mathcal{L}\) between model spectra \(M[\lambda]\) and UCD candidate NIR-spectra \(C[\lambda]\),
 \[\mathcal{L}\propto -\frac{1}{2} \sum_{\{\lambda\}} w[\lambda]\left[ \frac{C[\lambda] - \alpha M[\lambda]}{\sigma_C[\lambda]} \right]^2,\]
 where \(\alpha\) is left as a free parameter. Spectral regions used for the fit are the same as for Sect. \ref{sec:spec_class_method}. The parameter space is explored through a Markov Chain Monte Carlo (MCMC) approach, relying on the implementation of the \textit{emcee Python} package \citep{foreman-mackey_emcee_2013} and assuming uniform priors.

\begin{figure*}
     \centering
     \includegraphics[width=\linewidth]{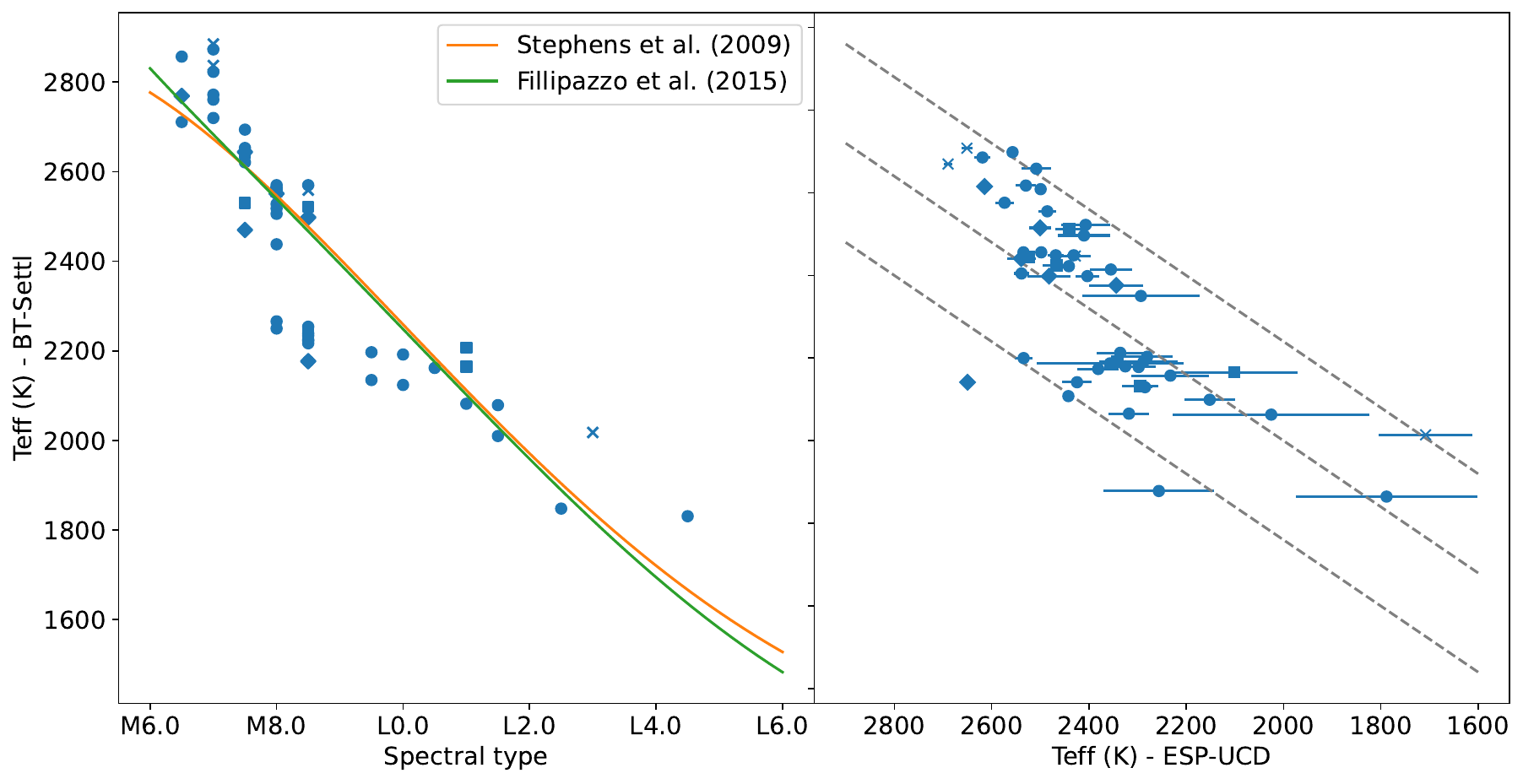}
     \caption{\textit{}Comparison between the temperatures (blue) against the spectral types found in this work (left). The \cite{stephens_08-145_2009} and \cite{filippazzo_fundamental_2015} spectral-type -- \(T_\mathrm{eff}\) relations are also shown in orange and green. \textit{}Comparison between the temperatures found in this work and those of the \gaia ESP-UCD pipeline (right).  The identity line is plotted in grey as well as the identity line \(\pm 300\) K. Symbols are the same as those used in Fig. \ref{fig:sofiGrpCAMD}.}
     \label{fig:teff-stephens_UCD}
 \end{figure*}

Model spectra are linearly interpolated from the most recent BT-Settl CIFIST grid \citep{allard_models_2012, allard_progress_2013, allard_synthetic_2014, baraffe_new_2015}. This new grid\footnote{Not yet made available to the public.}, provided by D. Homeier (priv. comm.), computed with the PHOENIX atmosphere code \citep{hauschildt_parallel_1997, allard_limiting_2001} and using solar abundances from \cite{caffau_solar_2011}, permits to determine the atmospheric parameters of UCDs. It was previously used at higher temperatures, to characterise properties, and particularly the metallicities, of M dwarfs \citep{hejazi_chemical_2020, hejazi_chemical_2022, zhang_m_2023}. In this work, we focus on the determination of the effective temperatures. We fixed the surface gravity at \(\log g = 4.5 \, \mathrm{[dex]}\), metallicity and alpha-enrichment at solar abundance (\([Fe/H] = 0\), \([\alpha/Fe] = 0\)), and interpolate the models only in effective temperatures, that are spaced by 100~K between 1500 and 2900~K. The surface gravity and the metallicity of objects will be studied using various atmosphere models in a future work.
 On the left panel of Fig. \ref{fig:teff-stephens_UCD}, we show their temperature, determined from their near-infrared spectra,  as function of the near-infrared spectral types. 
 Put in comparison with the relations derived by \cite{stephens_08-145_2009} and \cite{filippazzo_fundamental_2015} that associate UCD spectral types to temperatures and that are very close in this spectral type range, we observe that these temperatures are in agreement with the ones expected from the two relations. 
 Moreover, 55 objects observed in our sample have effective temperatures determined by the Extended Stellar Parametrizer for UCDs ESP-UCD \citep{sarro_ultracool_2023}.
The right panel of Fig. \ref{fig:teff-stephens_UCD} shows the difference between the ESP-UCD temperatures and our determinations. The scattering between the two temperatures is about 200 K. The hottest UCDs have a higher effective temperature than the estimation obtained from the optical \gaia observations (ESP-UCD). We provide the derived temperatures on Table \ref{tab:sofiobservations}, obtained from the posterior median, together with the standard deviation of the sample, as the posterior is typically Gaussian (an example is shown in the Appendix \ref{app:corner}).

\section{Binaries and peculiar cases}
\label{sec:peculiars_and_bins}

 In the sample, we have found objects that exhibit certain peculiarities. Here, we review objects that are best-matched with peculiar spectra, as well as objects showing signs of unresolved binarity. We also identified objects in wide binaries in the sample, which we highlight in this section. 
We searched for non-single sources in our sample in the \gaia non-single-stars tables \citep{halbwachs_gaia_2023}, and found none. In addition, we checked whether any of the observed UCDs could be members of nearby young clusters or associations. Using the online BANYAN \(\Sigma\) tool \citep{gagne_banyan_2018}, a Bayesian analysis tool that compares the kinematics of objects with the one of such structures, we found that none of our objects have been found to be members of such structures.

\subsection{Suspected subdwarfs}
\label{sec:potsubd}

Four objects in our sample appear blue or are well-matched by a subdwarf spectrum. These objects are suspected to have a low-metallicity. We attempted to recover their radial velocities using the Na~I doublet (11~385~\AA, 11~410~\AA) and the K~I doublets (11~692~\AA, 11~778~\AA ~and 12~437~\AA, 12~529~\AA), without success due to the low resolution of spectra.

\begin{figure}
     \centering
     \includegraphics[width=\linewidth]{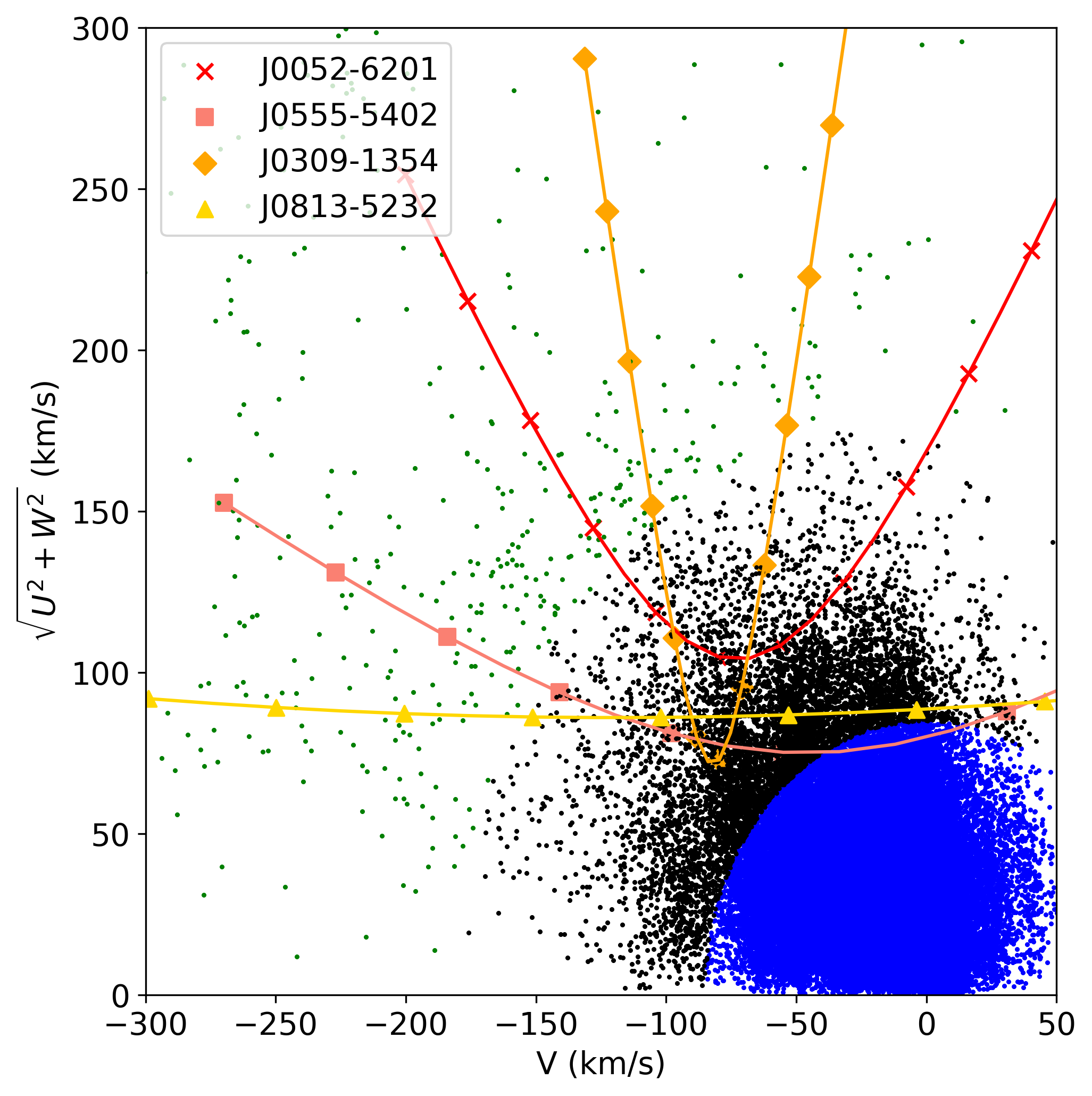}
     \caption{Position on Toomre diagram of the four potential subdwarfs (see text), assuming different radial velocities (in red:\ markers are spread out every 50 km/s, varies from -300 to 300 km/s). Background dots are sources from the \gaia Catalogue of Nearby Stars, with known radial velocities: halo stars are displayed in green, thick disk stars are in black, and thin disk stars are in blue. }
     \label{fig:toomre_diagrams}
 \end{figure}

In this section, we detail their peculiarities, and attempt to determine if they belong to an old population of stars. Similarly to \cite{kirkpatrick_field_2021}, we used \cite{nissen_thin_2004} kinematics criteria to separate fast-moving stars of the thick disk (with \(V_\mathrm{tot} > 85 \,\mathrm{km\,s^{-1}} \)) and of the halo (with \(V_\mathrm{tot} > 180\,\mathrm{km\,s^{-1}} \)) from slowly moving stars, which may belong to any population, from the thin disk to the halo. These kinematics criteria are indicative. A measurement of their metallicity and abundances of \(\alpha\)-elements would bring further constraints on the determination of their population.
Figure \ref{fig:toomre_diagrams} shows the Toomre diagrams of the four potential subdwarfs. As we do not have indication about their radial velocity, we allow it to range from \(-300\) to \(+300\,\mathrm{km\,s^{-1}}\) and visually assess whether they might be members of an old stellar population.
Their spectra are visible on Fig. \ref{fig:sofi-spec-pec}. 

 \begin{itemize}
     \item J0052-6201 resembles to a M7.0 subdwarf. It appears as slightly blue (\(G-G_{RP} = 1.435\pm 0.000\), \(J-Ks = 0.783\pm 0.048\)) without being remarkable relatively to objects in that part of the colour-absolute magnitude diagram. It has a high proper motion, with a tangential velocity of \(122\pm 0.1\,\mathrm{km\,s^{-1}}\), making it the fastest object in the sample, which could indicate an old age. We verify in its Toomre diagram that the object is likely a halo or a thick disk member, independently of its radial velocity, explaining its resemblance to a subdwarf. 

     \item J0555-5402 has a slightly higher absolute magnitude than J0052-6201 and resemble an object in between a dwarf and subdwarf (d/sd) M7.0. It is also slightly blue in Gaia bands, (\(G-G_{RP} = 1.435\pm 0.000\)) but not in 2MASS bands (\(J-Ks = 0.888\pm 0.059\)). It has a tangential velocity of \(79\pm 0.1\,\mathrm{km\,s^{-1}}\). Its Toomre diagram shows it might be a very fast thin disk member, but that depending on its radial velocity, it could belong to any population. As it is matched both by a d/sd and standard M7 templates, it may be either a dwarf or a subdwarf.

     \item  J0309-1354 is a particularly blue M.8.5, whose spectra match closely the NIR spectrum of WISEPC J010637.07+151852.8. This template object is analysed by \cite{kirkpatrick_first_2011} as having a slightly subsolar metallicity from its blue colour, and having high proper motion. Similarly, J0309-1354 is bluer in \(G-J\) colour relative to the rest of the sample (\(G-J = 4.21\pm 0.03\)) and exhibits large (projected) proper motions, with \(\mu = 0.7\pm 0.2\) arcsec/yr, corresponding to a tangential velocity of \(108\pm0.4\,\mathrm{km\,s^{-1}}\). Its kinematics are compatible with kinematics of the thick disk or halo stars.

     \item J0813-5232 is an L3 with high proper motions (\(\mu = 0.8\pm 0.4\) arcsec/yr, \(V_{tan} = 88\pm 0.7 \,\mathrm{km\,s^{-1}}\)). Its kinematics are the ones of a thick disk or halo member. It is closely matched by the spectrum of WISEA J071552.38-114532.9 from \cite{kirkpatrick_allwise_2014}, and appears bluer than standard L3 spectra. It is slightly blue in its J-Ks colour. 
 \end{itemize}

\subsection{Objects with peculiar spectra}
\label{sec:peculiars}

We found five UCDs with peculiar spectra that resemble previously published peculiar object spectra and/or  for which no standard or template spectra are found to match them visually. In this section, we detail their peculiarities, relative to other objects of similar spectral types:

\begin{itemize}

    \item J0325+1412 is a blue M.8.5, whose spectrum matches the NIR spectrum of WISEPC J010637.07+151852.8, an object with subsolar metallicity analysed by \cite{kirkpatrick_first_2011}. It has a rather low tangential velocity of \(34\pm 0.4\,\mathrm{km\,s^{-1}}\), typical of a disc star.

    \item J0552-0002 is a blue M9 that resembles to SIMP J22030176-0301107, a peculiar M9.5 observed by \cite{robert_brown_2016}. 

    \item J0827-5216 is found to be a peculiar L1. In Sect. \ref{sec:specbin}, we identify it as a spectral binary, explaining its peculiarity. 

    \item J0517-2816 is a blue L1, and its spectrum is similar to the one of J14403186-1303263 \citep{kirkpatrick_discoveries_2010}. Similarly to this source, J0517-2816 appears slightly bluer than a main-sequence star in the 2MASS J-Ks, H-Ks and J-H colours. It does not show peculiarities in \gaia colours.

    \item J0412-0734 was previously identified as a peculiar L2 by \cite{kirkpatrick_field_2021}. It appears particularly blue in 2MASS J-Ks colour (\(J-K_s = 0.32\pm0.03\)). It has relatively high proper motions, with \(\mu = 0.59\pm0.03\) arcsec/yr. However, this does not convert to a high tangential velocity (\(47\pm0.2 \,\mathrm{km\,s^{-1}}\)), the star being at 16 pc from the Sun.
    It is well-matched by the spectrum of SIMP J1811556+272840, a peculiar L2.5 observed by \cite{robert_brown_2016}.

\end{itemize}

\subsection{Common parallaxes and proper motions objects}
\label{sec:commonprop}

We observed a set of UCDs that we found to be part of a binary system from common parallaxes and proper motions in \gaia DR2 \citep{smart_gaia_2019, reyle_ultra-cool_2021}, where companions to UCD candidates are identified through following criteria : 
\[\rho < 100 \varpi, \]
\[\Delta\varpi < \max[1.0, 3\sigma_\varpi],\]
\[\Delta\mu<0.1\mu,\]
\[\Delta\theta<15^{\circ}.\]
Here \(\rho\) is the query radius, corresponding to a maximum projected separation of 100 000 AU from the source star, which is a conservative upper limit according to \cite{caballero_reaching_2009}. Also, \(\Delta\varpi\) is the parallax range in the line-of-sight around the UCD candidate, \(\Delta\mu\) is the difference in proper motion, and \(\Delta\theta\) is the angular difference between the proper motions vectors.

The seven binary systems components proper motions, parallaxes, and spectral types are listed in Table \ref{tab:bincomponents}. We use CoMover \citep{gagne_comover_2021}, which uses parallaxes, proper motions, and sky position of sources to assess the probability that objects are indeed in common proper motions and gravitationally bound. All systems are found to have a 99\% co-moving probability.

\begin{figure}
    \centering
    \includegraphics[width=\linewidth]{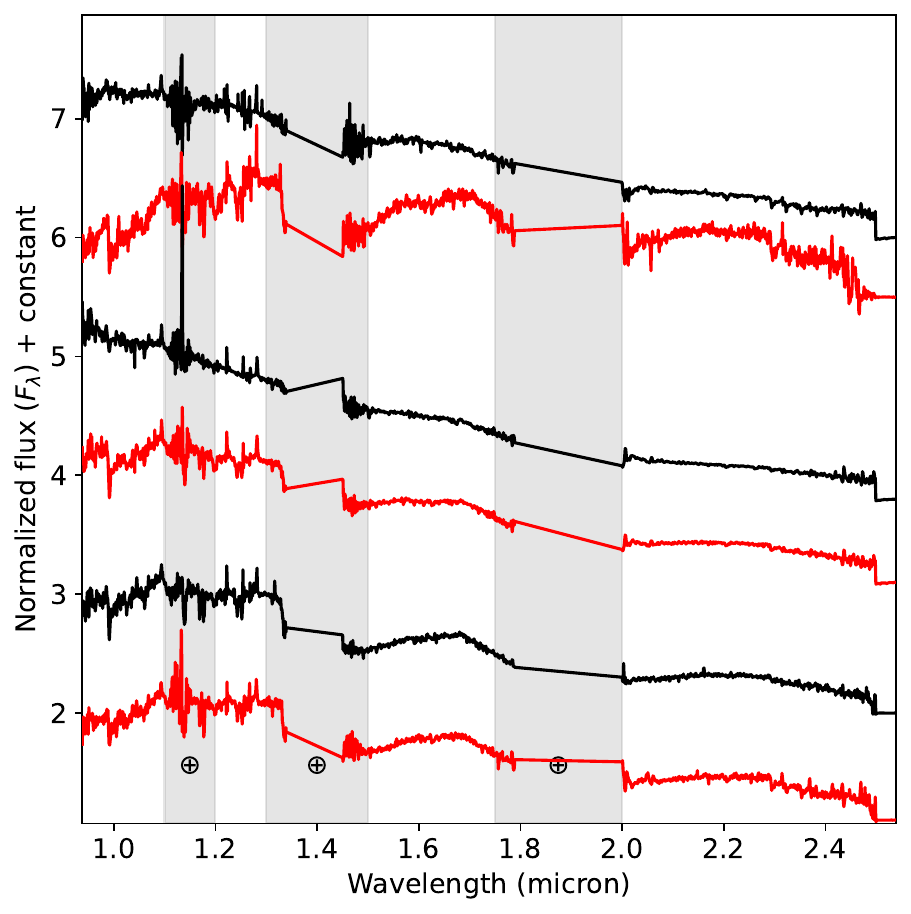}
    \caption{Binary system spectra. The primaries are shown in black, the secondaries in red. From top to bottom: 2MASS J23495194-2627594 (M5), J2349-2627 (L2.5); 2MASS J07215434-3104365 (M3), J0721-3105 (M6.5); and J0943-2009 (M7.5), J0943-2010 (M8.5).}
    \label{fig:binaries}
\end{figure}

For three systems, we were able to have both components on the same field of view and oriented the slit to acquire the spectra of the second component of the systems (see Fig. \ref{fig:binaries}). For the remaining systems, we only acquired the spectra of the UCD companion. An additional binary system, composed of J0347+0417 and Gaia 3271777035212786944, is found in Gaia DR3 but not DR2, due to a lack of parallax for the second object. These objects are not resolved by the ground spectrometer, as they are less distant than \(0.75''\). The observed spectrum shows signs of spectral binarity (see Sect. \ref{sec:specbin}).

\begin{table*}
    \centering
    \begin{tabular}{l l l l l l l}
        Name & Spectral Type & RA & Dec & \(\mu_{RA}\) & \(\mu_{DEC}\) & Parallax \\
         & & deg & deg & $\mathrm{mas\,yr^{-1}}$ & $\mathrm{mas\,yr^{-1}}$ & mas \\
         \hline\hline
         2MASS J23495194-2627594\tablefootmark{a} & M5.0 & 357.4676880 & -26.4661806 & 237.744 & 72.853 & 23.274 \\
         J2349-2627 & L2.5 & 357.4723326 & -26.4647437 & 237.085 & 73.130 & 21.905\\
         \hline
         2MASS 07215434-3104365 & M3.0 & 110.4767231 & -31.0778963 & 52.475 & -227.870 & 41.1380 \\
         J0721-3105 & M6.5 & 110.4802981 & -31.0885526 & 57.029 & -229.768 & 41.048  \\
         \hline
         J0943-2009 & M7.5 & 145.8666910 & -20.1648161 & -221.041 & 180.053 & 19.937 \\
         J0943-2010 & M8.5 & 145.8661622 & -20.1676840 & -219.376 & 179.819 & 19.932 \\
         \hline
         Gaia DR2 4757030391786232576 & WD & 82.3906350 & -63.9487136 & 166.109 & -4.879 & 14.048 \\
         J0529-6357 & M7.5 & 82.4196342 & -63.9525915 & 165.976 & -3.968 & 14.122\\
         \hline
         J0945-4120  & M8.0 & 146.3288956 & -41.3450184 & -144.533 & 83.965 & 17.139 \\
         2MASS 09450332-4115209 & M5? & 146.2629519 & -41.2554272 & -145.015 & 84.615 & 17.357 \\
         \hline
         J1456-5059 & M8.5 & 224.1588143 & -50.9860720 & -419.117 & -128.584 & 31.110 \\
         2MASS 14563857-5059174  & M3? & 224.1576894 & -50.9887596 & -420.868 & -146.202 & 30.799\\
         \hline
         J0347+0417 & M8 & 56.7820854 & 4.2981726 & 113.323 & -49.331 & 30.928 \tablefootmark{b} \\
         Gaia DR3 3271777035212786944 & M9.5? & 56.7820761 & 4.2983795 & 121.010 & -45.425 & 31.00 \\
         \hline         
    \end{tabular}
    \caption{Binary systems found from common parallaxes and proper motions, with at least a UCD candidate in their components. ''?'' denotes an unconfirmed spectral type, issued from photometric relations of \cite{reyle_new_2018} or \cite{hawley_characterization_2002}}
    \label{tab:bincomponents}
\tablefoot{
\tablefoottext{a}{identified in the RAVE survey \citep[RAVE DR6,][]{steinmetz_sixth_2020} as a \(0.10 M_\odot\), 9 Gyr star, with a radial velocity of \(12.00\pm3.85 \,\mathrm{km\,s^{-1}}\).}
\tablefoottext{b}{Analysed as a spectral binary, see Sect. \ref{sec:specbin}.}

}
\end{table*}

We find that J0529-6357 (2MASS J05294026-6357091) is a M7.5 companion to the white-dwarf \gaia DR2 4757030391786232576 spotted by \gaia, from its clear position in the white-dwarf sequence on the colour-absolute magnitude diagram. However, it is not possible to disentangle it from being a DA or a DB white dwarf from its locus on the colour-absolute magnitude diagram. Its total age can be obtained similarly to what was done by \cite{lodieu_3d_2019}. Using \cite{tremblay_improved_2011, bergeron_comprehensive_2011, blouin_new_2018} and \cite{bedard_spectral_2020} white-dwarfs cooling models, the cooling-age of the white dwarf can be obtained. The initial-to-final mass relation of \cite{el-badry_empirical_2018} permits to obtain its progenitor mass, and its age is found using the Padova evolutionary models \citep{bressan_parsec_2012}. Thus, we find that the white dwarf could be a DA white dwarf with a total age of 7.7 Gyr. Using evolutionary models from \cite{baraffe_new_2015}, we find that the ultra-cool dwarf, at such ages, should be a 0.08-0.095 \(M_\odot\) object, lying in the stellar regime.

    \subsection{Spectral binaries}
    \label{sec:specbin}

    Spectral binaries are multiple systems not resolved through photometric surveys. They can be revealed through typical features in their NIR spectra which are signatures of an unseen T-dwarf companion \citep{burgasser_spex_2010}. This method has been proved to be efficient from the detection of such a companion with HST high resolution imaging \citep{burgasser_hubble_2011}. To detect the typical spectral binaries features, \cite{burgasser_spex_2010, bardalez_gagliuffi_spex_2014} developed a set of 13 indices with 12 relations between them. They allow for the spectra of single stars to be distinguished from M/L+T dwarf spectral binaries. The indices are described in Table 3 of \cite{bardalez_gagliuffi_spex_2014} and originate from the shape of the J, H, and K bands and of the behaviour of different molecules (\(H_2 O, CH_4\)) in UCD atmospheres. The 12 relations permit to define regions on indice- indice diagrams, in which spectral binaries are more often located. However, a spectral binary will not be seen as such by all indices-indices relations, and single UCD spectra might also verify some of these relations. \cite{bardalez_gagliuffi_spex_2014} defined a spectrum as showing strong signs of being a from binary when eight or more relations are verified, weak signs if between four and eight relations are verified; otherwise,  the spectrum is likely to come from a non-binary. The spectra of objects detected as binary candidates have to be visually checked : as discussed by \cite{bardalez_gagliuffi_spex_2014}, this method can produce erroneous results for blue objects, which can be mistaken for binaries.
    
\begin{figure}
        \centering
        \includegraphics[width=\linewidth]{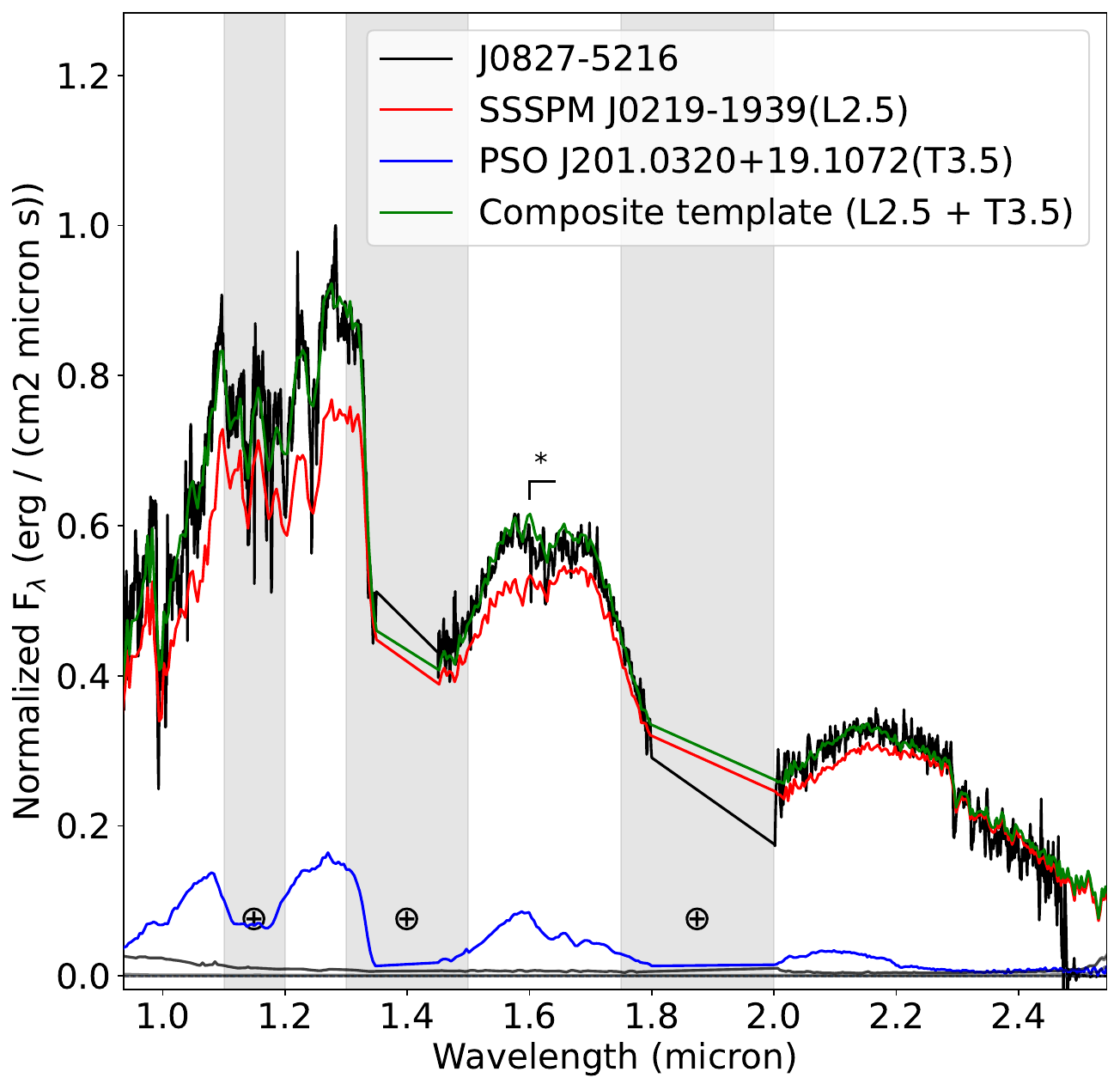}
        \caption{Spectrum of J0827-5216 and template spectra of its component, a L2.5 (red) and a T3.5 (blue), as well as their composite template (green). The H-dip caused by methane absorption of the T dwarf atmosphere is indicated by the star symbol.} 
        \label{fig:strongbin}
    \end{figure}

    We report that J0827-5216 exhibits strong signs of being a spectral binary, selected by 11 out of 12 indices. Using the template-fitting method described in \cite{bardalez_gagliuffi_spex_2014}, we find that it might be a L2.5 + T3.5 binary, as shown in Fig. \ref{fig:strongbin}. Its H-band notably show a dip, caused by \(CH_4\) absorption in the T-dwarf atmosphere, which is not present in the atmosphere of brown dwarfs with earlier spectral types. 

    We also find that 17 objects in our sample exhibit weak characteristics of M/L + T binaries. Following \cite{bardalez_gagliuffi_spex_2014}, we computed the F-test statistic between binary and single templates matching the spectra. This statistic represents the confidence level that the binary template is likely to be a better match than the single one. \cite{burgasser_spex_2010} considered that the confidence level, computed with an F-test to reject the null hypothesis, should be \(>99\%\) for an object to be a spectral binary, while \cite{bardalez_gagliuffi_spex_2014} chose it to be \(>90\%\) for being a binary candidate, complemented with a visual inspection. 
    In Table \ref{tab:Potbin}, we list seven objects that are potential spectral binaries, with an F-test score greater than 90\%. We also show the matching spectral types for the primaries, secondaries, as well as the best matching single spectral type. Their spectra are available in the appendix. 
    
    Among these objects, we report that J0347+0417 (which is in a binary system with \gaia DR3 3271777035212786944) is flagged as a potential spectral binary due to the presence of CH\(_4\) in the spectra. It might be composed of a M7.5 and a T2.0 according to our analysis. Using the \(M_G\)-spectral type relation of \cite{reyle_new_2018}, we find that the two \gaia resolved components should have spectral types of M7.5 and M9.5. Thus, J0347+0417A (M7.5), \gaia DR3 3271777035212786944 (M9.5?), and J0347+0417B (T2.0) might form a triple-UCD system. 

    We also find that J1906-0515 (\gaia DR3 4206320171755704320), identified as a M9 + T3.5, has a large RUWE in \gaia DR3 (2.48) and a positive \textit{IPD frac multi peak} (1). Moreover, its parallax varied between DR2 and DR3 (from 49.12 mas to 54.64 mas). These are strong signs of unresolved binarity, making the object a robust spectral binary candidate.

    \begin{table*}[!hbt]
        \centering
        \begin{tabular}{lllllll}
        Shortname & Gaia DR3 ID & F-test & Primary & Secondary & Single & Score\\
        \hline\hline
J0924-6655 & 5247231368807380224 & 0.99 & M8 & T3.5 & M7 & weak \\
J0010-0746 & 2429635550212644608 & 0.96 & M7.5 & T2.0 & M7.5 & weak\\
J0006+0439 & 2741783362285662592 & 0.98 & M8 & T1.5 & M8 & weak\\
J0347+0417 & 3271777035211085312 & 0.91 & M7.5 & T2.0 & M8 & weak\\
J1150-2850 & 3480933043353667328 & 0.96 & M9.5 & T4 & M9 & weak\\
J1906-0515 & 4206320171755704320 & 0.99 & M9: & T3.5 & M9.5 & weak\\
J0827-5216 & 5321416174249027584 & 0.99 & L2.5 & T3.5 & L2 & strong \\
\hline

        \end{tabular}
        \caption{Potential spectral binaries and their spectral types}
        \label{tab:Potbin}
    \end{table*}

\section{Conclusions}
We obtained the NIR spectra and spectroscopic spectral types of 60 objects selected from the catalogues of \cite{reyle_new_2018, smart_gaia_2019, scholz_new_2020}, using the SOFI spectrometer. These UCD candidates are closer than 50 pc or part of binary systems with common parallaxes and proper motions. Using SPLAT \citep{burgasser_spex_2017} to analyse their near-infrared spectra, a template-matching method is applied to confirm that 57 candidates are UCDs. The remaining three are photometric M7s that we find to be M6.5s. We thus complemented the local census of objects within 50 pc with new M to L dwarf spectra.

Using the precise astrometry and photometry of \gaia, UCD candidates can be selected through their absolute magnitude and can be assigned photometric spectral types. In this work, we show that \(M_G\)-spectral types of \cite{reyle_new_2018} found for the observed UCD candidates are similar to our spectroscopic types down to 1 subtype. 
 We find the use of colours-spectral type relations offer worse results, with photometric spectral types that have a precision of three subtypes compared to the spectroscopic ones.
This illustrates that \gaia observations can lead to an accurate estimation of the number of UCDs observed by the satellite. We are thus confident about the nature of the thousands of UCD candidates selected using \gaia magnitudes and parallaxes. 

We find that six objects we observed have already published spectral types. These are close to the ones we obtain, particularly if they were observed in the near-infrared. If the published spectral types were issued from the optical range, they slightly differ from our determination, as spectra acquired in these bands do not probe the same regions of the UCDs atmospheres.

We also carried out follow-up analyses of objects having common parallaxes and proper motions in our sample, and provide spectra of UCD candidates companions. We are able to constrain the mass and age of J0529-6357 thanks to its white dwarf binary companion, \gaia DR2 4757030391786232576.

Additionally, we retrieved the indices of spectral binaries in our sample. We found a strong binary candidate, 2MASS J08270052-5216277, which we expect to be a L2.5+T3.5. We also retrieve six objects with weaker indices of being spectral binaries, but whose composite template binary spectra are in better agreement with the observed one than with single templates. One of the objects, J0347+0417, is potentially a triple-UCD system. 


The UCD spectra and spectral types will be used in forthcoming studies. They can be compared with various atmosphere models covering the stellar-substellar transition effective temperatures (\(T_\mathrm{eff}<2700K\)) to derive their stellar parameters. These derivations will improve our knowledge about the atmospheric properties at the stellar-substellar transition and can be used to detail the wavelength region where the atmosphere modelling must be improved.

This sample is part of a greater survey that has the objective to complete the local census of nearby UCDs. Within 30 parsecs, 328 M7-M9.5 and 236 L0-L5.5 UCDs have been spectroscopically confirmed  \citep[extension of the catalogue from ][]{smart_gaia_2017}. Although our analysis does not contribute significantly to the characterisation of L dwarfs, it increased by 15\% the number of confirmed M-type UCDs. Starting from the 20 parsec sample and applying a simple volume scaling, we estimated that about 22\% of M7-L5.5 UCDs are missing in the 30 parsecs sample and are still yet to be discovered, observed, or spectroscopically confirmed. This census can be used to constrain formation processes in the solar neighbourhood, such as the multiplicity of low-mass stars and UCDs, and the stellar mass distribution at the end of the main sequence. 

\begin{acknowledgements}
This work has made use of data from the European Space Agency (ESA) mission Gaia (https://www.cosmos.esa.int/gaia), processed by the Gaia Data Processing and Analysis Consortium (DPAC, https://www. cosmos.esa.int/web/gaia/dpac/consortium).
This research has made use of the SIMBAD database,
operated at CDS, Strasbourg, France.
Based on observations collected at the European Organisation for Astronomical Research in the Southern Hemisphere under ESO programmes 106.214E.001 and 108.22G4.001.
This research has been supported by the Centre National d'Etudes Spatiales (CNES) PhD grant 2021-262, and a PhD grant from the Région Bourgogne-Franche-Comté. 
Processing steps have been executed on computers from the Utinam Institute of the Université de Franche-Comté, supported by the Région de Franche-Comté and Institut des Sciences de l’Univers (INSU).
T.R., C.R., N.L. acknowledge financial support from the ``Programme National de Physique Stellaire'' (PNPS) and ``Programme National de Cosmologie et Galaxies'' (PNCG) of CNRS/INSU, France.
J.G.F-T gratefully acknowledges the grant support provided by Proyecto Fondecyt Iniciaci\'on No. 11220340, and also from the Joint Committee ESO-Government of Chile 2021 (ORP 023/2021).

\end{acknowledgements}

%
%
\bibliographystyle{aa}
\bibliography{references}

\clearpage
\onecolumn

\begin{appendix}

\section{Spectra of the UCD candidates}
\label{app:sofi-spectra}
\begin{figure*}[h]
    \centering
    \includegraphics[width=\linewidth]{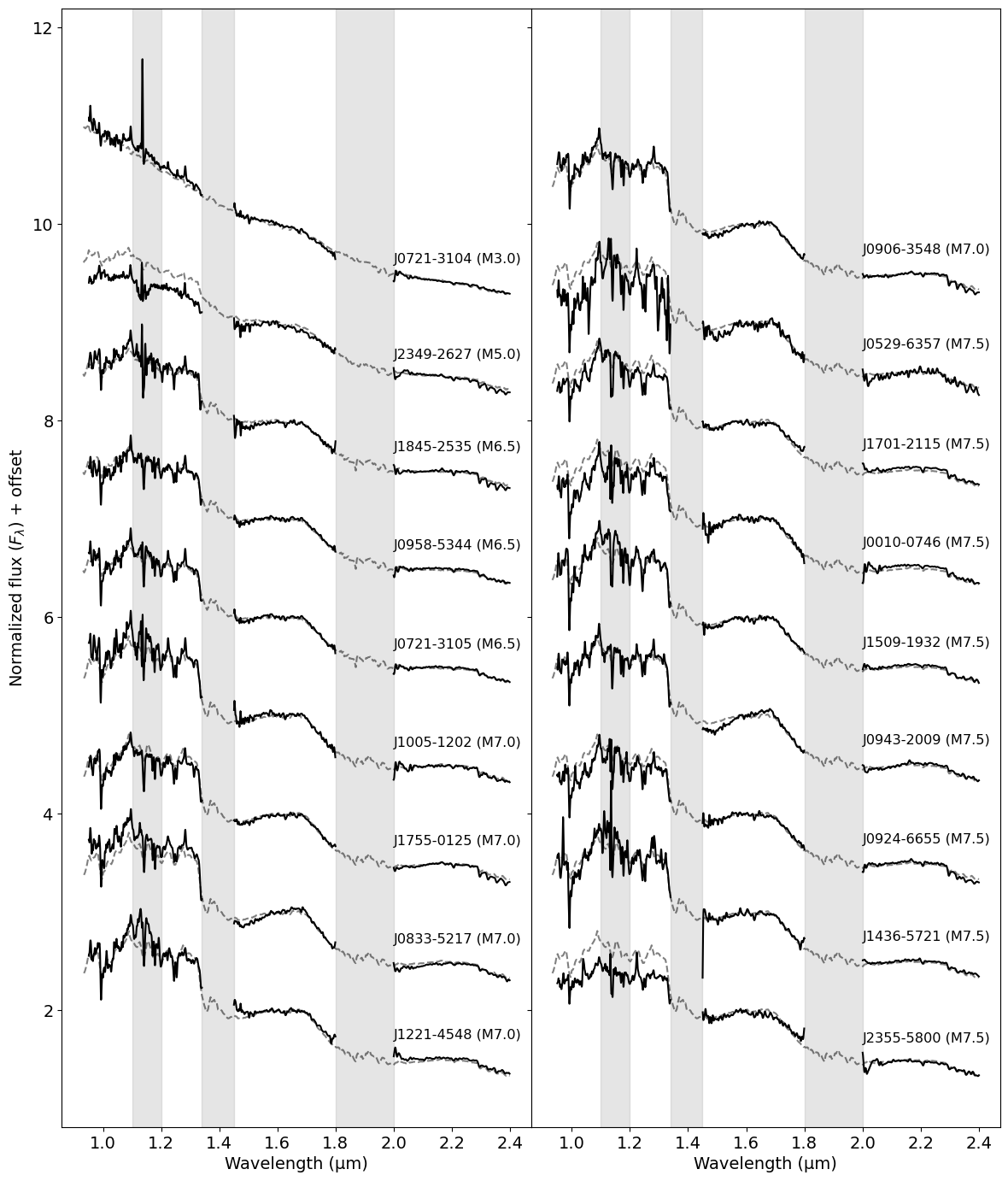}
    \caption{SOFI spectra (filled lines) of observed objects, together with standard spectra of the same spectral type (dashed lines, rounded to the nearest integer subtype). Object names and spectral types are annotated on the figure.}
    \label{fig:sofi-spec-1}
\end{figure*}

\begin{figure*}[h]
    \centering
    \includegraphics[width=\linewidth]{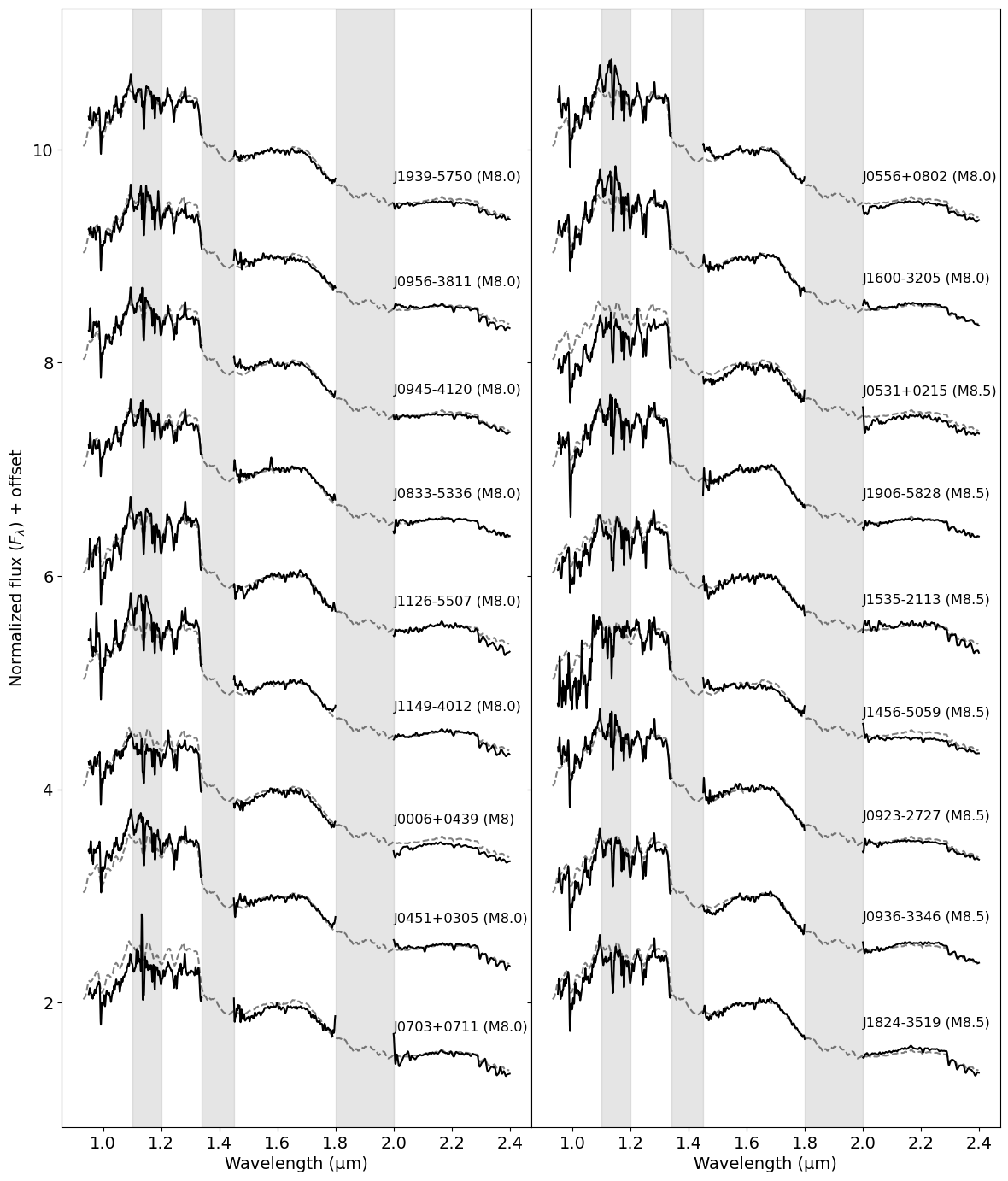}
    \caption{SOFI spectra (filled lines) of observed objects, together with standard spectra of the same spectral type (dashed lines, rounded to the nearest integer subtype). Object names and spectral types are annotated on the figure.}
    \label{fig:sofi-spec-2}
\end{figure*}

\begin{figure*}[h]
    \centering
    \includegraphics[width=\linewidth]{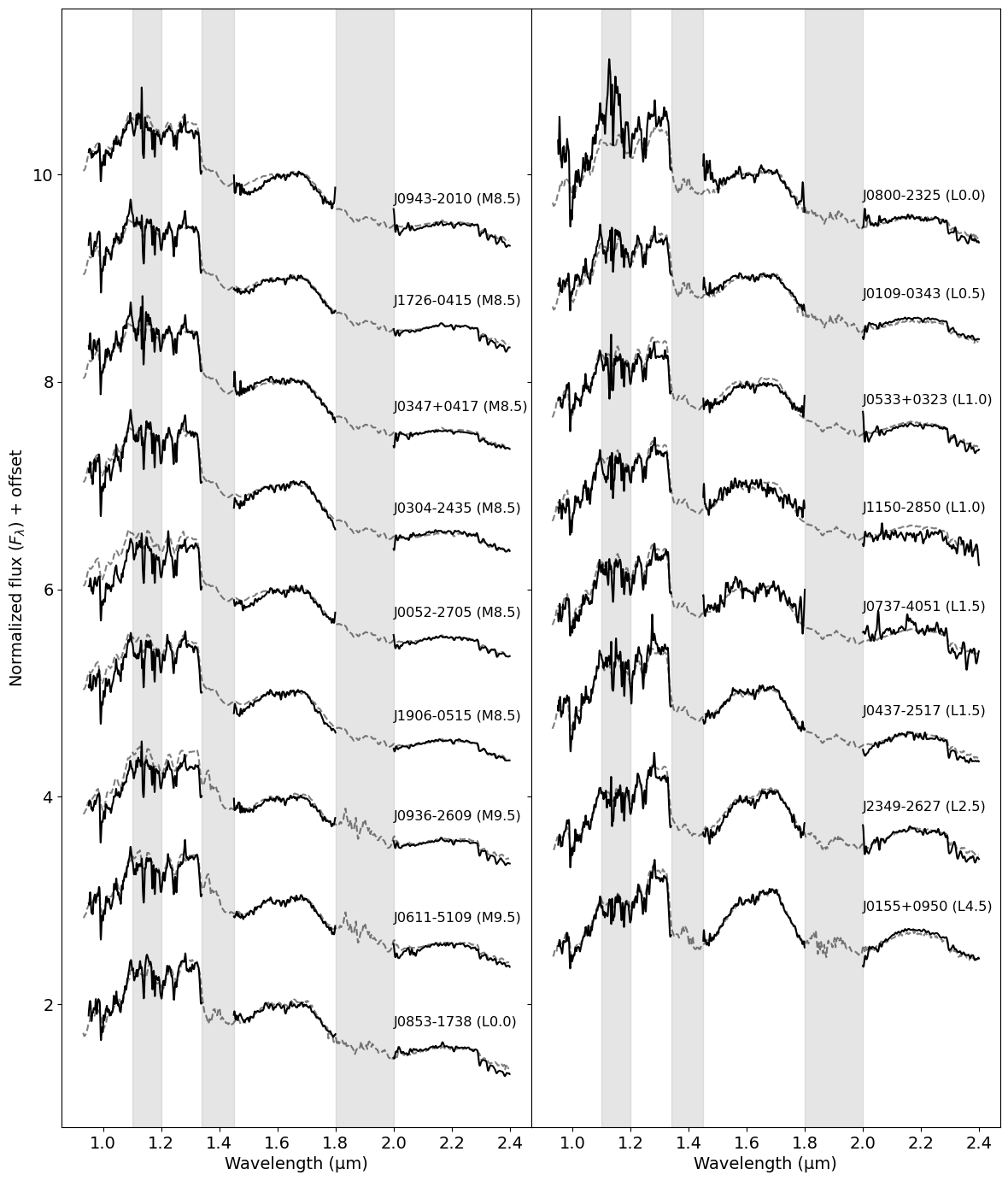}
    \caption{SOFI spectra (filled lines) of observed objects, together with standard spectra of the same spectral type (dashed lines, rounded to the nearest integer subtype). Object names and spectral types are annotated on the figure.}
    \label{fig:sofi-spec-3}
\end{figure*}

\begin{figure*}[h]
    \centering
    \includegraphics[width=\linewidth]{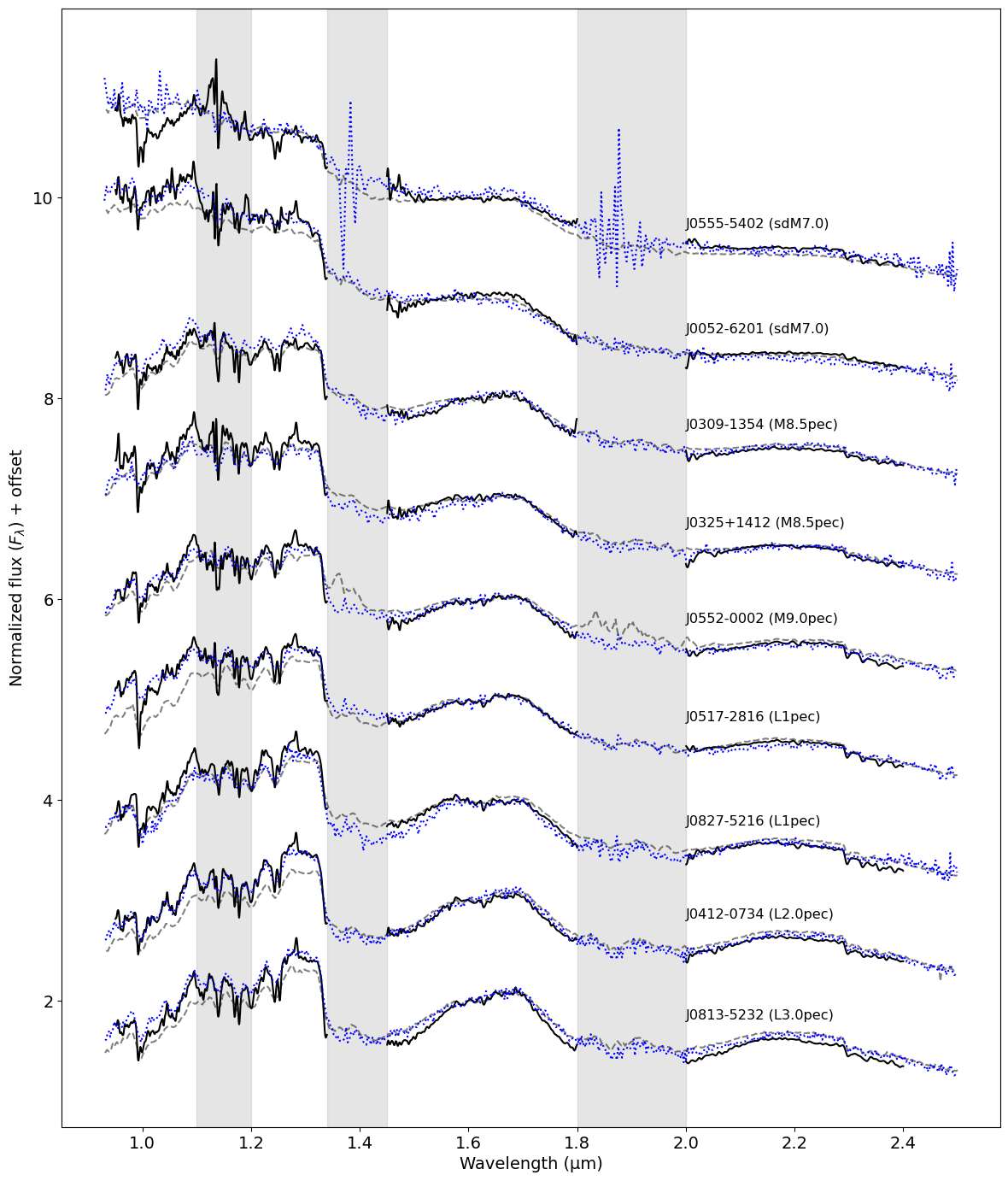}
    \caption{SOFI spectra (filled lines) of the peculiar objects of Sect. \ref{sec:peculiars_and_bins}, together with standard spectra of the same spectral type (dashed lines, rounded to the nearest integer subtype), and in dotted blue the best matching templates. Object names and spectral types are annotated on the figure.}
    \label{fig:sofi-spec-pec}
\end{figure*}

\clearpage

\section{Composite spectra of spectral binaries}

\label{app:composite}
   \begin{figure}[ht]
        \centering
        \includegraphics[width=0.5\linewidth]{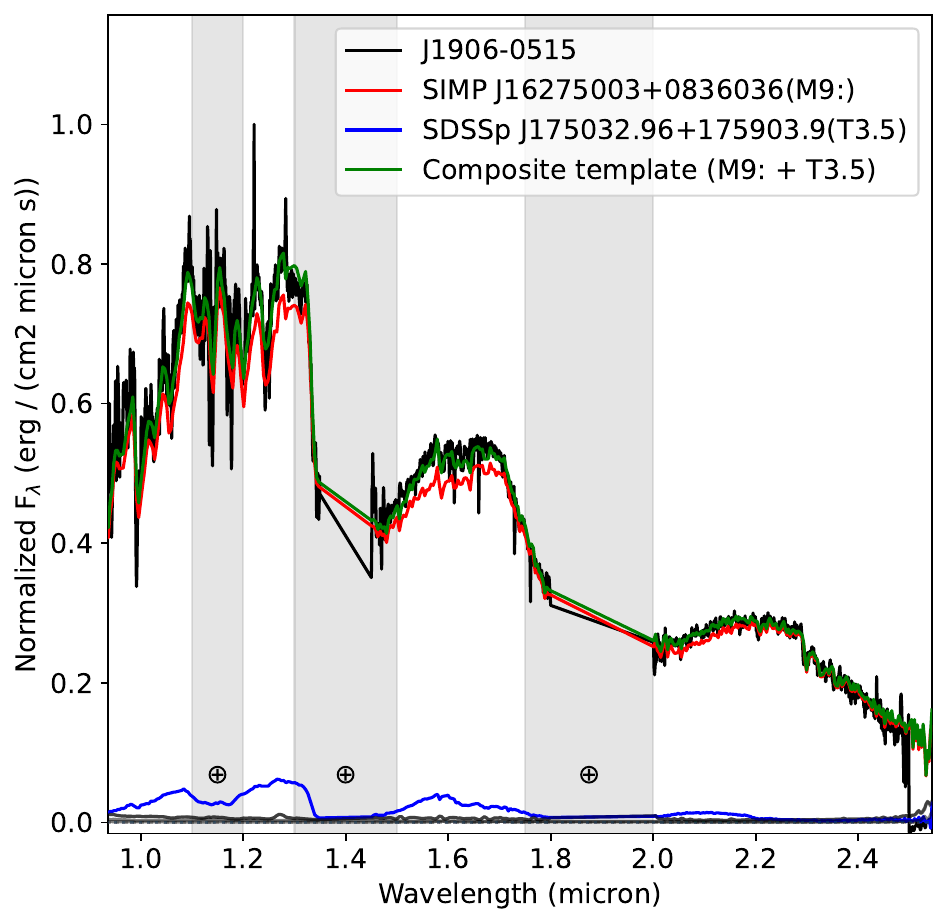}
        \caption{Spectrum of J1906-0515 and template spectra of its component, a M9 (red) and a T3.5 (blue), as well as their composite template (green).} 
    \end{figure}

       \begin{figure}[ht]
        \centering
        \includegraphics[width=0.5\linewidth]{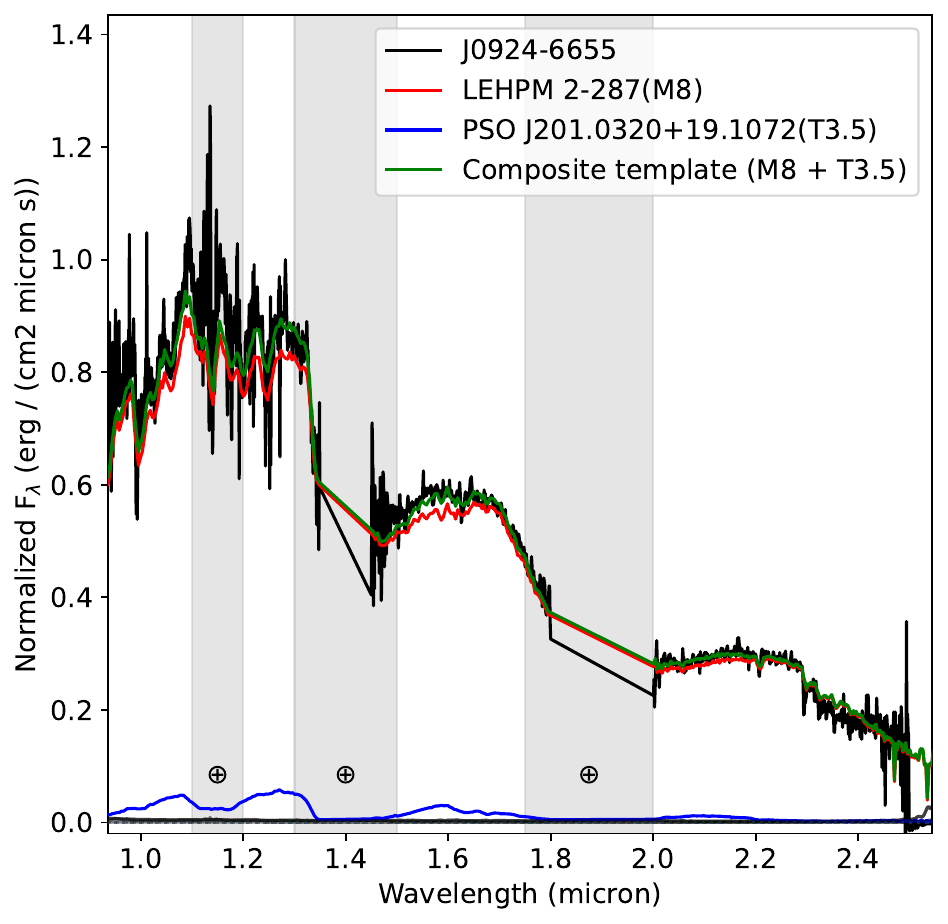}
        \caption{Spectrum of J0924-6655 and template spectra of its component, a M8 (red) and a T3.5 (blue), as well as their composite template (green).} 
    \end{figure}

       \begin{figure}[ht]
        \centering
        \includegraphics[width=0.5\linewidth]{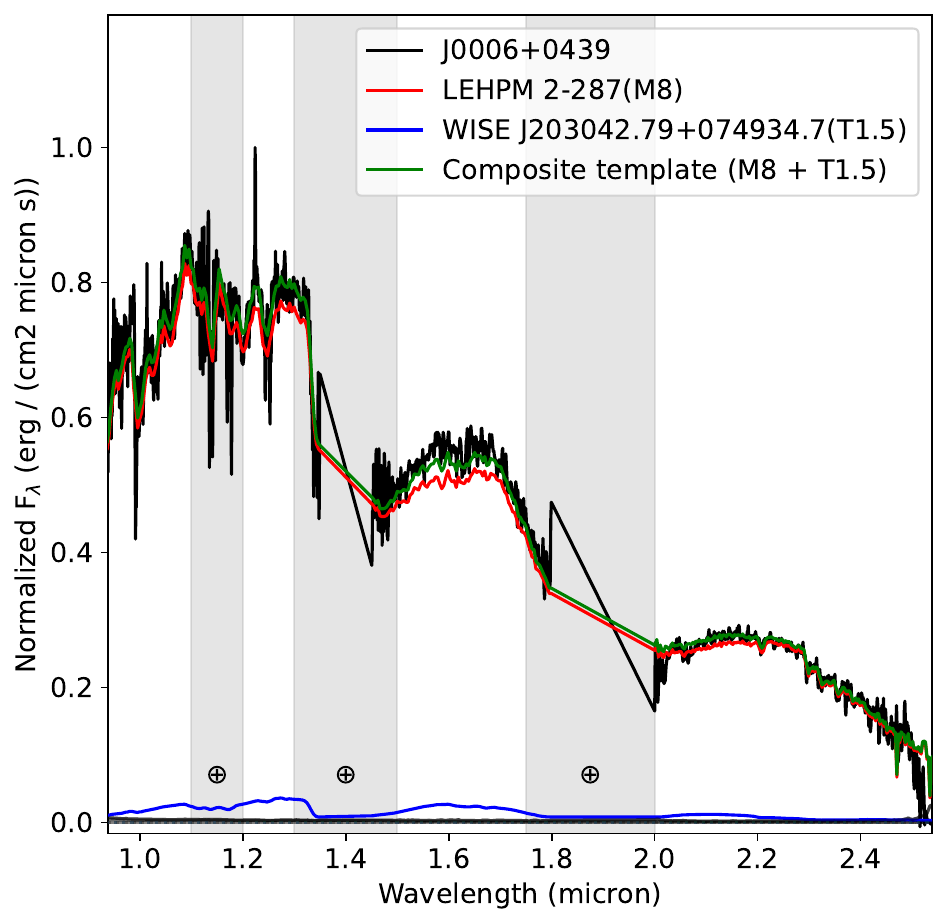}
        \caption{Spectrum of J0006+0439 and template spectra of its component, a M8 (red) and a T1 (blue), as well as their composite template (green).} 
    \end{figure}

    \begin{figure}[ht]
        \centering
        \includegraphics[width=0.5\linewidth]{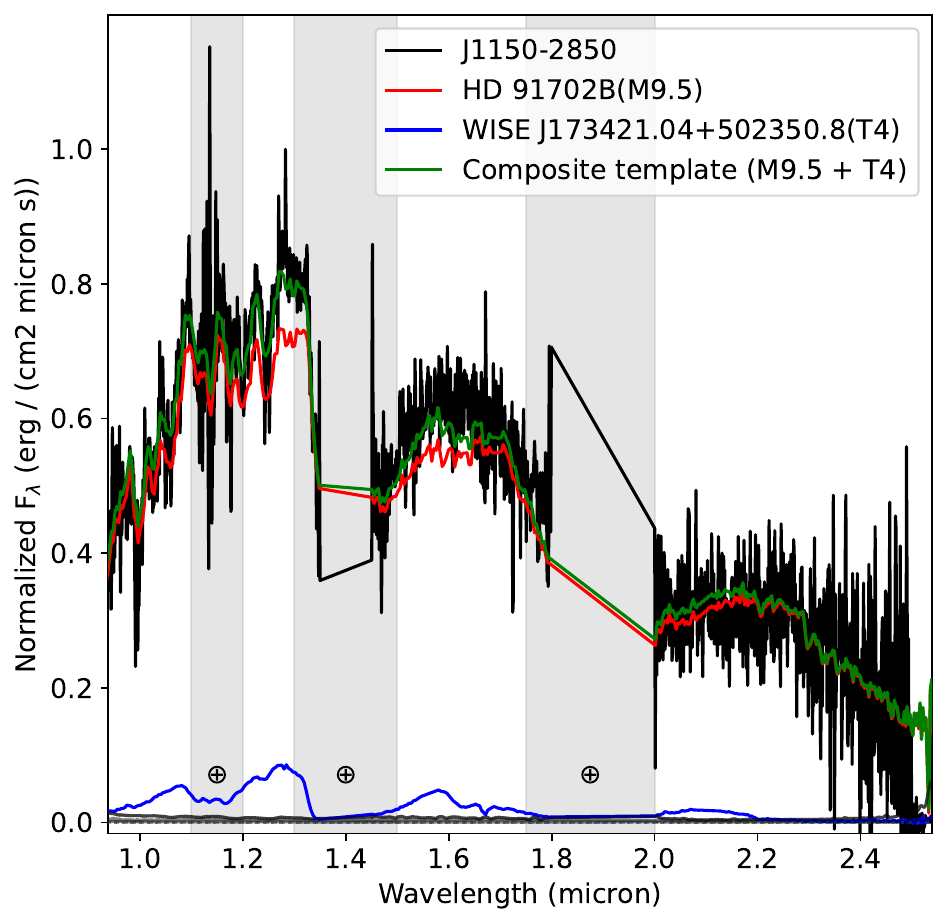}
        \caption{Spectrum of J1150-2850 and template spectra of its component, a M9.5 (red) and a T4 (blue), as well as their composite template (green)} 
    \end{figure}

   \begin{figure}[ht]
        \centering
        \includegraphics[width=0.5\linewidth]{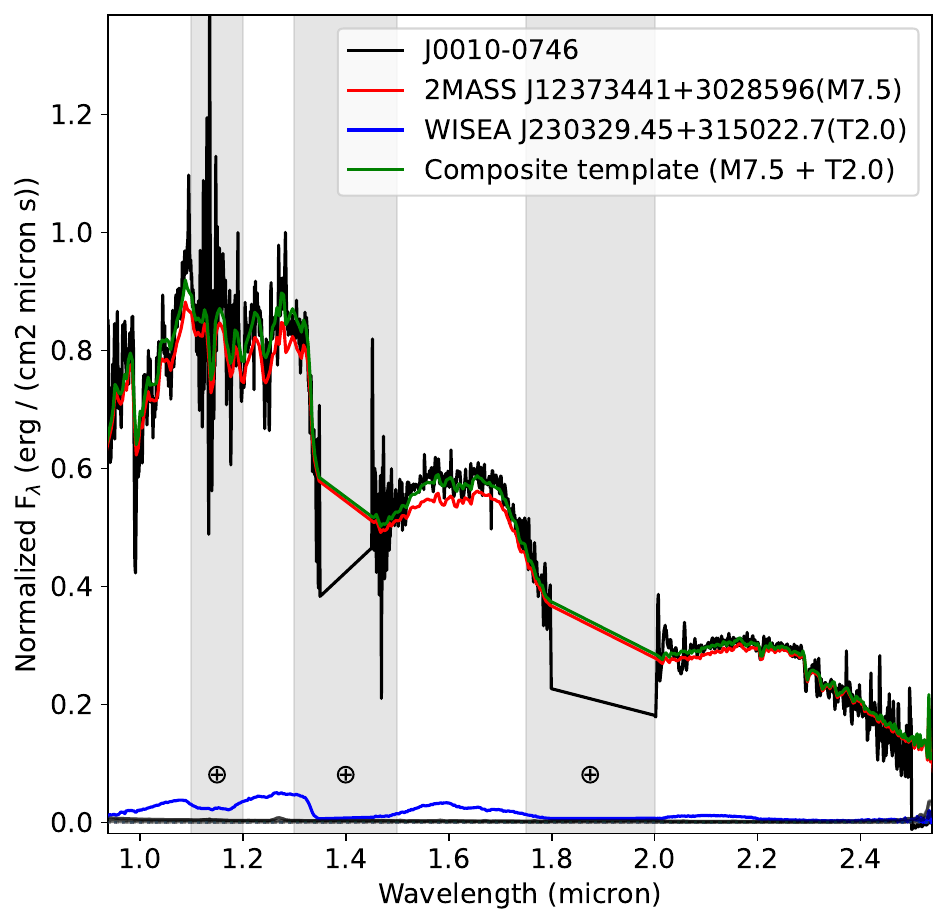}
        \caption{Spectrum of J0010-0746 and template spectra of its component, a M7.5 (red) and a T2 (blue), as well as their composite template (green).} 
    \end{figure}

   \begin{figure}[ht]
        \centering
        \includegraphics[width=0.5\linewidth]{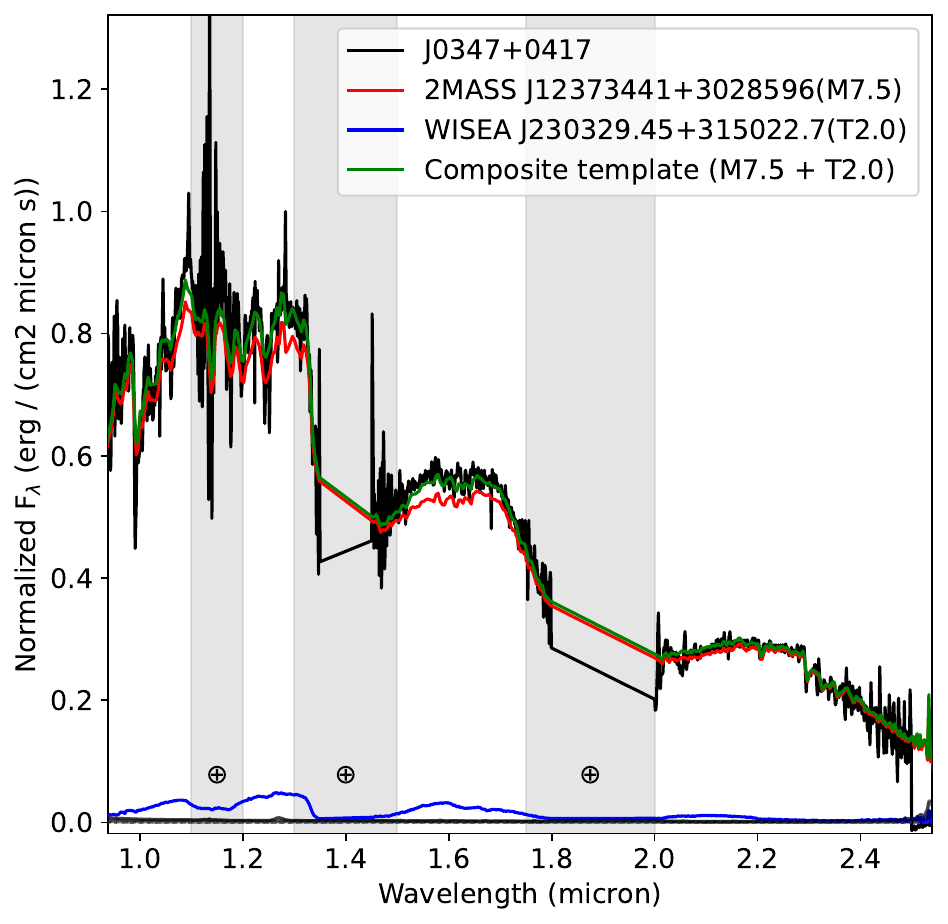}
        \caption{Spectrum of J0347+0417 and template spectra of its component, a M7.5 (red) and a T2 (blue), as well as their composite template (green).} 
    \end{figure}

\clearpage

\section{Corner plot}
\label{app:corner}
\begin{figure}[h]
    \centering
    \includegraphics[width=\linewidth]{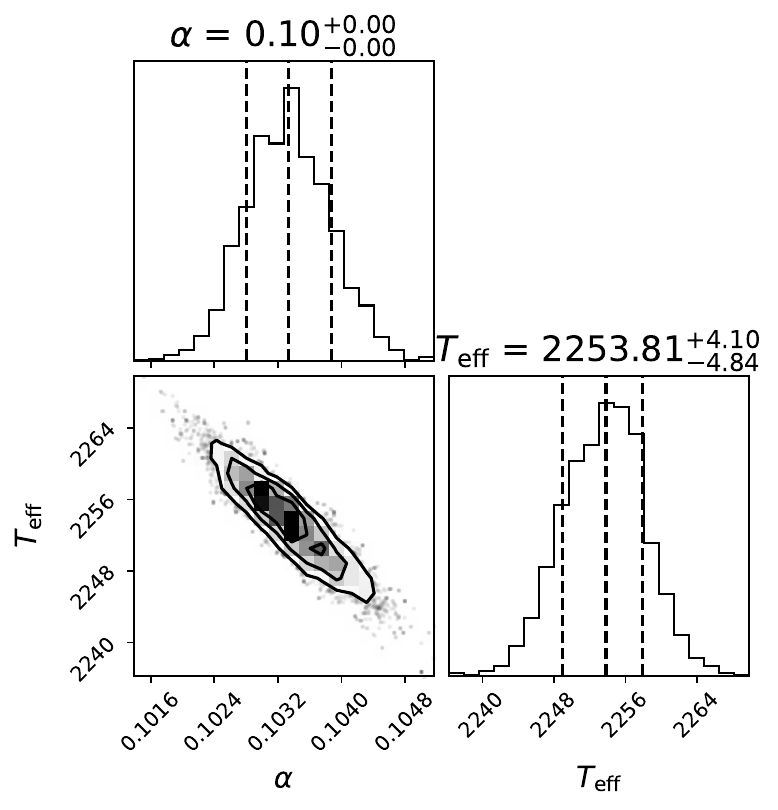}
    \caption{Corner plot from the fitting process of \(T_\mathrm{eff}\) and the scale parameter \(\alpha\) (Section \ref{sec:efftemp}), for J0304-2435.}
\end{figure}

\clearpage
\onecolumn
\section{Tables describing observed objects}

\setcounter{table}{0}

        \begin{longtable}{lllllllllll}
            \caption{Shortnames, RA, DEC, parallaxes \(\varpi\), proper motions, and tangential velocities of the UCDs candidates observed by SOFI.\\
            \label{tab:sofiobservations}}\\
            \hline\hline
            Shortname & RA & Dec & \(\varpi\) & \(\sigma_\varpi\) & \(\mu_\mathrm{RA}\) & \(\sigma_{\mu\mathrm{RA}}\) & \(\mu_\mathrm{DEC}\) & \(\sigma_{\mu\mathrm{DEC}}\) & \(V_\mathrm{tan}\) & \(\sigma_{V_\mathrm{tan}}\)\\
             & HMS & DMS & mas & mas & $\mathrm{mas\,yr^{-1}}$ & $\mathrm{mas\,yr^{-1}}$ & $\mathrm{mas\,yr^{-1}}$ & $\mathrm{mas\,yr^{-1}}$ &  \(\mathrm{km\,s^{-1}}\) & \(\mathrm{km\,s^{-1}}\)\\
            \hline
            \endfirsthead
            \caption{continued.}\\
            \hline\hline
             Shortname & RA & Dec & \(\varpi\) & \(\sigma_\varpi\) & \(\mu_\mathrm{RA}\) & \(\sigma_{\mu\mathrm{RA}}\) & \(\mu_\mathrm{DEC}\) & \(\sigma_{\mu\mathrm{DEC}}\) & \(V_\mathrm{tan}\) & \(\sigma_{V_\mathrm{tan}}\)\\
            \hline
            \endhead
            \hline
            \endfoot
J0721-3104 \tablefootmark{a} & 07:21:54.4 & -31:04:40.4 & 41.138 & 0.016 & 52.475 & 0.016 & -227.87 & 0.018 & 26.94 & 0.01 \\
J2349-2627 \tablefootmark{a}  & 23:49:52.2 & -26:27:58.3 & 23.274 & 0.018 & 237.744 & 0.016 & 72.853 & 0.016 & 50.63 & 0.04 \\
J0958-5344 & 09:58:42.4 & -53:44:51.0 & 46.074 & 0.039 & 242.005 & 0.043 & -198.693 & 0.039 & 32.21 & 0.03 \\
J0721-3105 & 07:21:55.3 & -31:05:18.8 & 41.048 & 0.052 & 57.029 & 0.048 & -229.768 & 0.061 & 27.34 & 0.03 \\
J1845-2535 & 18:45:08.0 & -25:35:18.1 & 34.849 & 0.091 & 200.831 & 0.092 & -302.312 & 0.074 & 49.37 & 0.15 \\
J0906-3548 & 09:06:42.0 & -35:48:20.5 & 31.964 & 0.095 & -143.549 & 0.083 & 69.212 & 0.097 & 23.64 & 0.06 \\
J1221-4548 & 12:21:48.7 & -45:48:41.6 & 30.154 & 0.129 & -286.954 & 0.102 & 29.2 & 0.088 & 45.35 & 0.16 \\
J0555-5402 & 05:55:15.1 & -54:03:02.2 & 38.368 & 0.052 & 84.333 & 0.063 & -631.34 & 0.068 & 78.68 & 0.11 \\
J1755-0125 & 17:55:28.5 & -01:26:03.4 & 34.649 & 0.115 & -139.829 & 0.136 & -210.216 & 0.11 & 34.55 & 0.08 \\
J0833-5217 & 08:33:59.2 & -52:17:17.3 & 26.505 & 0.089 & -171.595 & 0.101 & 461.856 & 0.116 & 88.07 & 0.29 \\
J1005-1202 & 10:05:16.6 & -12:02:10.3 & 34.674 & 0.13 & -105.083 & 0.134 & -16.061 & 0.129 & 14.53 & 0.03 \\
J0052-6201 & 00:52:17.8 & -62:01:52.3 & 42.464 & 0.043 & 1085.154 & 0.058 & 132.136 & 0.057 & 122.03 & 0.11 \\
J0529-6357 & 05:29:40.7 & -63:57:09.3 & 14.122 & 0.261 & 165.976 & 0.352 & -3.968 & 0.342 & 55.79 & 1.1 \\
J1701-2115 & 17:01:07.1 & -21:15:18.5 & 30.25 & 0.152 & -107.039 & 0.181 & -186.655 & 0.129 & 33.73 & 0.14 \\
J0924-6655 & 09:24:12.9 & -66:55:28.8 & 32.221 & 0.098 & -242.823 & 0.13 & 89.626 & 0.12 & 38.07 & 0.1 \\
J2355-5800 & 23:55:22.6 & -58:00:04.3 & 35.522 & 0.049 & 389.49 & 0.043 & -45.059 & 0.051 & 52.32 & 0.08 \\
J1436-5721 & 14:36:52.5 & -57:21:34.9 & 32.943 & 0.114 & 137.747 & 0.094 & -94.4 & 0.103 & 24.02 & 0.08 \\
J0010-0746 & 00:10:29.5 & -07:46:50.7 & 34.18 & 0.12 & 94.293 & 0.133 & -110.301 & 0.089 & 20.13 & 0.08 \\
J1509-1932 & 15:09:24.1 & -19:32:07.2 & 28.512 & 0.201 & -392.17 & 0.274 & -75.065 & 0.216 & 66.39 & 0.44 \\
J0943-2009 & 09:43:28.0 & -20:09:53.3 & 19.937 & 0.16 & -221.041 & 0.175 & 180.053 & 0.153 & 67.81 & 0.53 \\
J0703+0711 & 07:03:03.4 & +07:10:54.7 & 37.558 & 0.113 & 137.498 & 0.098 & -385.495 & 0.087 & 51.66 & 0.15 \\
J1149-4012 & 11:49:48.9 & -40:12:46.1 & 29.898 & 0.182 & -565.921 & 0.132 & -164.379 & 0.134 & 93.44 & 0.53 \\
J0451+0305 & 04:51:14.3 & +03:05:29.0 & 39.969 & 0.109 & 259.296 & 0.119 & 29.608 & 0.099 & 30.95 & 0.1 \\
J1600-3205 & 16:00:38.5 & -32:05:26.6 & 29.86 & 0.278 & -24.77 & 0.342 & -108.207 & 0.246 & 17.63 & 0.11 \\
J1126-5507 & 11:26:31.9 & -55:07:42.5 & 20.123 & 0.407 & 10.845 & 0.407 & -19.914 & 0.369 & 5.35 & 0.08 \\
J0006+0439 & 00:06:26.6 & +04:39:07.9 & 36.617 & 0.137 & 322.529 & 0.174 & -234.084 & 0.145 & 51.59 & 0.2 \\
J0945-4120 & 09:45:18.9 & -41:20:42.1 & 17.139 & 0.155 & -144.533 & 0.136 & 83.965 & 0.152 & 46.26 & 0.41 \\
J1939-5750 & 19:39:52.2 & -57:50:32.0 & 46.897 & 0.094 & 75.261 & 0.077 & 114.179 & 0.078 & 13.82 & 0.04 \\
J0833-5336 & 08:33:42.7 & -53:36:35.0 & 50.379 & 0.049 & -256.163 & 0.058 & 389.039 & 0.057 & 43.83 & 0.05 \\
J0556+0802 & 05:56:08.5 & +08:02:37.7 & 36.799 & 0.13 & 60.049 & 0.125 & 4.262 & 0.106 & 7.75 & 0.04 \\
J0956-3811 & 09:56:03.5 & -38:11:45.1 & 32.566 & 0.106 & 6.642 & 0.081 & -225.363 & 0.096 & 32.82 & 0.08 \\
J0531+0215 & 05:31:27.7 & +02:15:49.8 & 31.713 & 0.219 & 23.661 & 0.22 & 8.861 & 0.185 & 3.78 & 0.07 \\
J0347+0417 & 03:47:07.7 & +04:17:53.4 & 30.928 & 0.154 & 113.323 & 0.155 & -49.331 & 0.117 & 18.94 & 0.11 \\
J1726-0415 & 17:26:21.7 & -04:15:34.5 & 32.536 & 0.153 & 37.6 & 0.167 & 74.172 & 0.126 & 12.12 & 0.09 \\
J0052-2705 & 00:52:54.8 & -27:05:58.4 & 40.187 & 0.182 & 50.53 & 0.179 & 73.232 & 0.191 & 10.49 & 0.08 \\
J0304-2435 & 03:04:32.2 & -24:35:14.0 & 33.907 & 0.22 & -39.33 & 0.15 & -21.287 & 0.231 & 6.26 & 0.01 \\
J0309-1354 & 03:09:33.6 & -13:54:35.5 & 32.403 & 0.161 & 11.62 & 0.177 & -737.651 & 0.172 & 107.91 & 0.43 \\
J1906-5828 & 19:06:25.0 & -58:28:26.1 & 37.519 & 0.129 & -156.359 & 0.106 & -24.023 & 0.097 & 19.99 & 0.06 \\
J1906-0515 & 19:06:07.2 & -05:15:05.1 & 54.644 & 0.481 & 216.472 & 0.384 & -44.787 & 0.31 & 19.2 & 0.21 \\
J0325+1412 & 03:25:29.0 & +14:12:30.4 & 25.084 & 0.337 & -69.65 & 0.356 & -170.888 & 0.253 & 34.89 & 0.39 \\
J1535-2113 & 15:35:27.5 & -21:13:52.7 & 14.274 & 1.038 & -36.533 & 1.302 & -30.915 & 1.181 & 16.18 & 0.47 \\
J0923-2727 & 09:23:30.5 & -27:27:20.9 & 38.949 & 0.105 & 193.512 & 0.091 & -4.305 & 0.091 & 23.55 & 0.08 \\
J0943-2010 & 09:43:27.9 & -20:10:03.7 & 19.932 & 0.179 & -219.376 & 0.192 & 179.819 & 0.166 & 67.46 & 0.59 \\
J1456-5059 & 14:56:38.1 & -50:59:09.9 & 31.11 & 0.085 & -419.117 & 0.087 & -128.584 & 0.081 & 66.79 & 0.17 \\
J0936-3346 & 09:36:54.5 & -33:46:20.3 & 35.503 & 0.16 & -299.51 & 0.109 & 57.356 & 0.123 & 40.72 & 0.16 \\
J1824-3519 & 18:24:16.5 & -35:19:21.6 & 38.657 & 0.16 & 20.969 & 0.193 & 33.923 & 0.156 & 4.89 & 0.05 \\
J0552-0002 & 05:52:31.5 & -00:02:11.2 & 39.222 & 0.208 & 181.64 & 0.234 & 77.984 & 0.203 & 23.9 & 0.17 \\
J0611-5109 & 06:11:27.6 & -51:09:09.1 & 33.908 & 0.162 & 139.103 & 0.235 & 273.832 & 0.181 & 42.95 & 0.23 \\
J0936-2609 & 09:36:55.7 & -26:09:42.7 & 54.171 & 0.078 & 36.571 & 0.077 & -27.256 & 0.048 & 3.99 & 0.01 \\
J0853-1738 & 08:53:11.7 & -17:38:56.0 & 24.847 & 0.436 & -40.884 & 0.486 & -21.488 & 0.336 & 8.81 & 0.05 \\
J0800-2325 & 08:00:48.7 & -23:25:58.7 & 21.605 & 1.082 & -39.999 & 0.789 & -13.226 & 1.279 & 9.35 & 0.21 \\
J0109-0343 & 01:09:51.6 & -03:43:26.3 & 94.623 & 0.203 & 372.168 & 0.265 & 8.652 & 0.156 & 18.65 & 0.05 \\
J0517-2816 & 05:17:58.7 & -28:16:10.9 & 38.085 & 0.145 & 196.696 & 0.132 & -144.576 & 0.171 & 30.39 & 0.12 \\
J0533+0323 & 05:33:32.3 & +03:23:15.9 & 24.919 & 0.226 & -56.113 & 0.21 & -99.393 & 0.171 & 21.74 & 0.15 \\
J1150-2850 & 11:50:37.4 & -28:50:16.6 & 29.584 & 0.208 & -454.167 & 0.215 & -214.067 & 0.147 & 80.44 & 0.46 \\
J0827-5216 & 08:27:00.7 & -52:16:26.8 & 26.04 & 0.31 & 88.203 & 0.354 & 50.441 & 0.385 & 18.51 & 0.31 \\
J0437-2517 & 04:37:31.7 & -25:17:46.2 & 22.13 & 0.57 & -153.603 & 0.483 & -23.751 & 0.591 & 33.4 & 0.7 \\
J0737-4051 & 07:37:00.9 & -40:52:00.6 & 24.932 & 0.61 & -60.572 & 0.681 & -85.39 & 0.713 & 19.93 & 0.28 \\
J0412-0734 & 04:12:46.9 & -07:34:17.0 & 59.99 & 0.241 & 408.44 & 0.246 & -429.43 & 0.191 & 46.83 & 0.17 \\
J2349-2627 & 23:49:53.4 & -26:27:53.1 & 21.905 & 1.129 & 237.085 & 0.813 & 73.13 & 0.738 & 54.26 & 2.89 \\
J0813-5232 & 08:13:19.4 & -52:31:50.2 & 43.054 & 0.342 & -63.867 & 0.458 & 801.787 & 0.463 & 88.6 & 0.73 \\
J0155+0950 & 01:55:03.9 & +09:49:59.0 & 45.042 & 0.545 & 326.965 & 0.608 & -88.075 & 0.518 & 35.65 & 0.5 \\
\end{longtable}
\tablefoot{
\tablefoottext{a}{Not an UCD candidate, observed as in a binary system with a UCD candidate.}
}

  \begin{landscape}
        \begin{longtable}{lllllllllllllll}
            \caption{List of shortnames, \gaia DR3 IDs, \gaia and 2MASS magnitudes, spectral type and temperatures of the UCD candidates observed with SOFI.\\
            \label{tab:sofispectral}}\\
            \hline\hline
            Shortname & \gaia DR3 ID & \(G\) & \(\sigma_G\) & \(G_{RP}\) & \(\sigma_{G_{RP}}\) & \(J\) & \(\sigma_J\) & \(H\) & \(\sigma_H\) & \(Ks\) & \(\sigma_{Ks}\) & SpT & \(T_\mathrm{eff}\) (K) & \(\sigma_T\)  \\
            \hline
            \endfirsthead
            \caption{continued.}\\
            \hline\hline
             Shortname & \gaia DR3 ID & \(G\) & \(\sigma_G\) & \(G_{RP}\) & \(\sigma_{G_{RP}}\) & \(J\) & \(\sigma_J\) & \(H\) & \(\sigma_H\) & \(Ks\) & \(\sigma_{Ks}\) & SpT & \(T_\mathrm{eff}\) (K) & \(\sigma_T\)  \\
            \hline
            \endhead
            \hline
            \endfoot
J0721-3104 & 5604989633324007168 & 12.03 & 0.0 & 10.88 & 0.0 & 9.37 & 0.03 & 8.76 & 0.06 & 8.54 & 0.02 & M3.0 &  &  \\
J2349-2627 & 2337434628275058688 & 13.06 & 0.0 & 11.95 & 0.0 & 10.47 & 0.03 & 9.85 & 0.02 & 9.64 & 0.02 & M5.0 &  &  \\
J1845-2535 & 4073379904492310656 & 16.54 & 0.0 & 15.07 & 0.0 & 12.61 & 0.02 & 11.99 & 0.03 & 11.62 & 0.02 & M6.5 & 2860 & 19 \\
J0958-5344 & 5404351271207811968 & 15.63 & 0.0 & 14.2 & 0.0 & 11.89 & 0.03 & 11.29 & 0.02 & 10.99 & 0.02 & M6.5 & 2710 & 13 \\
J0721-3105 & 5604989328392927232 & 16.3 & 0.0 & 14.83 & 0.0 & 12.4 & 0.02 & 11.78 & 0.02 & 11.43 & 0.02 & M6.5 & 2770 & 22 \\
J1005-1202 & 3766333787676996736 & 17.27 & 0.0 & 15.73 & 0.0 & 13.07 & 0.03 & 12.43 & 0.02 & 12.11 & 0.03 & M7.0 & 2820 & 31 \\
J1755-0125 & 4178755415232881152 & 16.94 & 0.0 & 15.44 & 0.0 & 13.01 & 0.03 & 12.46 & 0.03 & 12.1 & 0.03 & M7.0 & 2720 & 19 \\
J0833-5217 & 5321275024453205632 & 17.44 & 0.0 & 15.93 & 0.0 & 13.58 & 0.03 & 13.01 & 0.03 & 12.68 & 0.03 & M7.0 & 2870 & 20 \\
J0555-5402 & 4767876769050721920 & 16.33 & 0.0 & 14.88 & 0.0 & 12.52 & 0.03 & 11.9 & 0.03 & 11.64 & 0.03 & sdM7.0 & 2880 & 13 \\
J1221-4548 & 6131567952556034304 & 17.54 & 0.0 & 16.01 & 0.0 & 13.35 & 0.03 & 12.71 & 0.03 & 12.3 & 0.03 & M7.0 & 2760 & 24 \\
J0906-3548 & 5624087604843625472 & 17.21 & 0.0 & 15.68 & 0.0 & 13.14 & 0.03 & 12.6 & 0.03 & 12.26 & 0.03 & M7.0 & 2770 & 25 \\
J0052-6201 & 4902366110781708288 & 15.85 & 0.0 & 14.41 & 0.0 & 12.15 & 0.02 & 11.74 & 0.03 & 11.37 & 0.02 & sdM7.0 & 2840 & 34 \\
J0529-6357 & 4757030327366948608 & 19.39 & 0.0 & 17.82 & 0.01 & 15.0 & 0.05 & 14.32 & 0.06 & 14.12 & 0.09 & M7.5 & 2470 & 12 \\
J1701-2115 & 4127833836291916800 & 17.94 & 0.0 & 16.4 & 0.0 & 13.54 & 0.03 & 12.83 & 0.03 & 12.37 & 0.03 & M7.5 & 2621 & 10 \\
J0010-0746 & 2429635550212644608 & 17.42 & 0.0 & 15.9 & 0.0 & 13.13 & 0.03 & 12.5 & 0.02 & 12.08 & 0.03 & M7.5 & 2530 & 11 \\
J1509-1932 & 6256820090941649920 & 18.19 & 0.0 & 16.65 & 0.01 & 13.9 & 0.03 & 13.19 & 0.02 & 12.82 & 0.03 & M7.5 & 2650 & 13 \\
J0943-2009 & 5665466110041259904 & 18.37 & 0.0 & 16.82 & 0.01 & 14.21 & 0.03 & 13.77 & 0.04 & 13.34 & 0.04 & M7.5 & 2640 & 12 \\
J0924-6655 & 5247231368807380224 & 17.68 & 0.0 & 16.15 & 0.0 & 13.32 & 0.02 & 12.67 & 0.02 & 12.25 & 0.02 & M7.5 & 2641 & 10 \\
J1436-5721 & 5891504088488622208 & 17.43 & 0.0 & 15.9 & 0.0 & 13.27 & 0.02 & 12.64 & 0.03 & 12.22 & 0.03 & M7.5 & 2690 & 22 \\
J2355-5800 & 6494861747014476288 & 16.14 & 0.0 & 14.71 & 0.0 & 12.39 & 0.02 & 11.8 & 0.03 & 11.42 & 0.02 & M7.5 & 2640 & 14 \\
J1939-5750 & 6448329418495400960 & 16.41 & 0.0 & 14.88 & 0.0 & 12.24 & 0.02 & 11.61 & 0.02 & 11.2 & 0.02 & M8.0 & 2570 & 15 \\
J0956-3811 & 5433620854830342272 & 17.54 & 0.0 & 16.0 & 0.0 & 13.14 & 0.02 & 12.42 & 0.02 & 12.01 & 0.03 & M8.0 & 2530 & 13 \\
J0945-4120 & 5431356479348370432 & 18.48 & 0.0 & 16.96 & 0.01 & 14.32 & 0.03 & 13.73 & 0.04 & 13.33 & 0.04 & M8.0 & 2550 & 13 \\
J0833-5336 & 5321049521489667328 & 16.11 & 0.0 & 14.61 & 0.0 & 12.02 & 0.02 & 11.35 & 0.02 & 10.98 & 0.02 & M8.0 & 2510 & 12 \\
J1126-5507 & 5346390614507242752 & 19.87 & 0.0 & 18.28 & 0.02 & 15.24 & 0.05 & 14.54 & 0.04 & 13.99 & 0.06 & M8.0 & 2440 & 13 \\
J1149-4012 & 5381471151464274816 & 18.54 & 0.0 & 16.99 & 0.01 & 14.0 & 0.03 & 13.32 & 0.02 & 12.83 & 0.02 & M8.0 & 2520 & 14 \\
J0006+0439 & 2741783362285662592 & 17.02 & 0.0 & 15.5 & 0.0 & 12.96 & 0.03 & 12.35 & 0.03 & 12.01 & 0.02 & M8 & 2560 & 12 \\
J0451+0305 & 3233120164884973824 & 17.0 & 0.0 & 15.45 & 0.0 & 12.73 & 0.02 & 12.06 & 0.02 & 11.7 & 0.02 & M8.0 & 2560 & 16 \\
J0703+0711 & 3153877399801800576 & 16.87 & 0.0 & 15.36 & 0.0 & 12.78 & 0.02 & 12.13 & 0.03 & 11.76 & 0.02 & M8.0 & 2250 & 6 \\
J0556+0802 & 3323397147633734016 & 17.47 & 0.0 & 15.94 & 0.01 & 13.16 & 0.03 & 12.5 & 0.03 & 12.1 & 0.02 & M8.0 & 2561 & 10 \\
J1600-3205 & 6036456382825238272 & 18.78 & 0.0 & 17.19 & 0.01 & 14.17 & 0.03 & 13.43 & 0.04 & 13.02 & 0.03 & M8.0 & 2270 & 14 \\
J0531+0215 & 3223589696879068416 & 18.79 & 0.0 & 17.2 & 0.01 & 14.16 & 0.03 & 13.47 & 0.03 & 13.02 & 0.04 & M8.5 & 2239 & 4 \\
J1906-5828 & 6633500847494138496 & 17.52 & 0.0 & 15.98 & 0.0 & 13.23 & 0.03 & 12.65 & 0.03 & 12.32 & 0.03 & M8.5 & 2177 & 8 \\
J1535-2113 & 6253313091230794368 & 20.67 & 0.01 & 19.08 & 0.05 & 16.05 & 0.08 & 15.15 & 0.07 & 14.9 & 0.1 & M8.5 & 2234 & 6 \\
J1456-5059 & 5901094750438455808 & 16.59 & 0.0 & 15.14 & 0.0 & 12.67 & 0.03 & 12.06 & 0.03 & 11.7 & 0.03 & M8.5 & 2177 & 4 \\
J0923-2727 & 5637175400984142336 & 17.02 & 0.0 & 15.49 & 0.0 & 12.89 & 0.03 & 12.24 & 0.02 & 11.91 & 0.02 & M8.5 & 2570 & 11 \\
J0936-3346 & 5437808478004461056 & 18.31 & 0.0 & 16.71 & 0.01 & 13.72 & 0.03 & 13.03 & 0.03 & 12.58 & 0.03 & M8.5 & 2225 & 4 \\
J1824-3519 & 6734417258848246400 & 17.74 & 0.0 & 16.16 & 0.0 & 13.26 & 0.02 & 12.55 & 0.02 & 12.11 & 0.02 & M8.5 & 2217 & 5 \\
J0943-2010 & 5665466110041260800 & 18.45 & 0.0 & 16.92 & 0.02 & 14.28 & 0.03 & 13.76 & 0.03 & 13.43 & 0.04 & M8.5 & 2500 & 11 \\
J1726-0415 & 4363458933311794816 & 18.03 & 0.0 & 16.45 & 0.0 & 13.6 & 0.02 & 12.93 & 0.03 & 12.56 & 0.02 & M8.5 & 2500 & 11 \\
J0347+0417 & 3271777035211085312 & 17.26 & 0.0 & 15.44 & 0.0 & 12.74 & 0.02 & 12.1 & 0.03 & 11.71 & 0.02 & M8.5 & 2520 & 11 \\
J0304-2435 & 5074762692133598848 & 18.71 & 0.0 & 17.12 & 0.01 & 14.08 & 0.02 & 13.34 & 0.03 & 12.89 & 0.03 & M8.5 & 2254 & 5 \\
J0309-1354 & 5156148920778713088 & 17.82 & 0.0 & 16.24 & 0.01 & 13.61 & 0.03 & 13.07 & 0.02 & 12.75 & 0.03 & M8.5 pec (blue) & 2560 & 11 \\
J0052-2705 & 2342686956666281728 & 18.19 & 0.0 & 16.61 & 0.01 & 13.61 & 0.03 & 12.98 & 0.03 & 12.54 & 0.03 & M8.5 & 2223 & 3 \\
J1906-0515 & 4206320171755704320 & 17.4 & 0.0 & 15.83 & 0.01 & 12.91 & 0.03 & 12.26 & 0.04 & 11.81 & 0.02 & M8.5 & 2239 & 5 \\
J0325+1412 & 41966709564899840 & 18.71 & 0.0 & 17.17 & 0.01 & 14.41 & 0.03 & 13.78 & 0.03 & 13.29 & 0.03 & M8.5 pec (blue) & 2478 & 10 \\
J0552-0002 & 3218721677865937664 & 18.33 & 0.0 & 16.72 & 0.01 & 13.71 & 0.02 & 13.11 & 0.03 & 12.64 & 0.02 & M9.0pec & 2237 & 3 \\
J0936-2609 & 5661194163772723072 & 16.55 & 0.0 & 15.03 & 0.0 & 12.26 & 0.03 & 11.61 & 0.03 & 11.2 & 0.02 & M9.5 & 2135 & 4 \\
J0611-5109 & 5549762268067194112 & 18.8 & 0.0 & 17.23 & 0.01 & 14.12 & 0.03 & 13.4 & 0.03 & 12.92 & 0.02 & M9.5 & 2197 & 6 \\
J0853-1738 & 5729782588170108800 & 19.67 & 0.0 & 18.13 & 0.01 & 14.92 & 0.04 & 14.11 & 0.03 & 13.53 & 0.05 & L0.0 & 2124 & 8 \\
J0800-2325 & 5699057186711474816 & 20.65 & 0.01 & 19.18 & 0.03 & 15.89 & 0.09 & 14.86 & 0.1 & 14.46 & 0.08 & L0.0 & 2190 & 12 \\
J0109-0343 & 2531195858721613056 & 16.38 & 0.0 & 14.8 & 0.0 & 11.69 & 0.02 & 10.93 & 0.03 & 10.43 & 0.02 & L0.5 & 2162 & 5 \\
J0533+0323 & 3224368700870633856 & 18.48 & 0.0 & 16.9 & 0.01 & 13.8 & 0.03 & 13.17 & 0.03 & 12.68 & 0.03 & L1.0 & 2082 & 6 \\
J0517-2816 & 2954622425142243840 & 18.13 & 0.0 & 16.54 & 0.01 & 13.6 & 0.03 & 13.01 & 0.03 & 12.58 & 0.02 & L1 pec (blue) & 2149 & 5 \\
J1150-2850 & 3480933043353667328 & 18.82 & 0.0 & 17.22 & 0.01 & 14.21 & 0.03 & 13.48 & 0.03 & 13.03 & 0.03 & L1.0 & 2160 & 11 \\
J0827-5216 & 5321416174249027584 & 19.7 & 0.0 & 18.09 & 0.02 & 14.82 & 0.07 & 14.08 & 0.08 & 13.5 & 0.04 & L1 pec (blue) & 2207 & 6 \\
J0737-4051 & 5536535589866627456 & 20.54 & 0.01 & 19.01 & 0.06 & 15.72 & 0.07 & 14.77 & 0.06 & 14.01 & 0.06 & L1.5 & 2010 & 30 \\
J0437-2517 & 4894335896327787392 & 20.37 & 0.0 & 18.82 & 0.03 & 15.43 & 0.06 & 14.77 & 0.06 & 14.41 & 0.08 & L1.5 & 2079 & 8 \\
J0412-0734 & 3195979005694112768 & 18.53 & 0.0 & 16.93 & 0.01 & 12.6 & 0.0 & 12.77 & 0.03 & 12.28 & 0.03 & L2.0 pec (blue) & 2068 & 7 \\
J2349-2627 & 2337434628274641024 & 20.21 & 0.01 & 18.69 & 0.04 & 15.36 & 0.06 & 14.61 & 0.06 & 14.1 & 0.07 & L2.5 & 1848 & 3 \\
J0813-5232 & 5320834841834013568 & 19.81 & 0.0 & 18.2 & 0.02 & 14.64 & 0.04 & 13.92 & 0.04 & 13.29 & 0.04 & L3.0pec & 2018 & 4 \\
J0155+0950 & 2570824594248255104 & 20.01 & 0.01 & 18.47 & 0.02 & 14.82 & 0.04 & 13.76 & 0.03 & 13.14 & 0.04 & L4.5 & 1831 & 2 \\
\end{longtable}
\end{landscape}

\end{appendix}

\end{document}